\documentclass[11pt,a4paper]{article}
\pdfoutput=1

\usepackage{cite}
\usepackage[english]{babel}
\usepackage[utf8]{inputenc}
\usepackage{amsmath}
\usepackage{graphicx}
\usepackage[nottoc]{tocbibind}
\usepackage{amssymb}

\usepackage[colorlinks=true, allcolors=blue]{hyperref}
\usepackage{float}
\usepackage{multirow}
\usepackage{makeidx}

\numberwithin{figure}{section}
\numberwithin{equation}{section}

\newcommand{\be}{\begin{equation}}
\newcommand{\ee}{\end{equation}}
\newcommand{\bea}{\begin{eqnarray}}
\newcommand{\eea}{\end{eqnarray}}
\def\beal#1\eeal{\begin{align}#1\end{align}}   
\def\besp#1\eesp{\begin{multline}#1\end{multline}} 

\newcommand{\TRM}[1]{#1}

\newcommand\ie{\textit{i.e.}\ }
\newcommand\eg{\textit{e.g.}\ }
\newcommand\cf{\textit{cf.}\ }

\newcommand{\etc}{{\it etc.}\ }
\newcommand{\viz}{{\it viz.}\ }

\newcommand{\nn}{\nonumber}

\newcommand{\ph}{\varphi}
\newcommand{\half}{\tfrac{1}{2}}

\newcommand{\sGamma}{\mathring{\Gamma}}
\newcommand{\Po}{\mathcal{P}}

\title{Ultraviolet finite resummation of perturbative quantum gravity}

\author{Tim R. Morris\\
\small \texttt{T.R.Morris@soton.ac.uk}\\
 \small \textit{STAG Research Centre, Department of Physics and Astronomy,}\\
  \small \textit{University of Southampton, Highfield, Southampton, SO17 1BJ, U.K.}
  }

\date{\today}
\makeindex
\begin{document}

\maketitle
\begin{abstract}
If the metric is chosen to depend exponentially on the conformal factor, and if one works in a gauge where the conformal factor has the wrong sign propagator, perturbative quantum gravity corrections can be partially resummed into a series of terms each of which is ultraviolet finite. These new terms however are not perturbative in some small parameter, and  are not individually BRST invariant, or background diffeomorphism invariant. 
With appropriate parametrisation, the finiteness property holds true also for a full phenomenologically relevant theory of quantum gravity coupled to (beyond the standard model) matter fields, provided massive tadpole corrections are set to zero by a trivial renormalisation. 
\end{abstract}
\newpage
\tableofcontents
\newpage

\section{Introduction}
\label{sec:Introduction}

As is well known, quantum gravity suffers from the problem that it is not perturbatively renormalizable.  Kinematic accidents allow pure gravity at one loop to be free of divergences \cite{tHooft:1974toh} (after a reparametrisation of the metric $g_{\mu\nu}$ off shell), but with generic matter or at two loops, no such miracle occurs \cite{tHooft:1974toh,Goroff:1985sz,Goroff:1985th,vandeVen:1991gw}. However, in order to make sense of the Feynman integrals, one works in Euclidean signature. Then, since the
Einstein-Hilbert action\footnote{We set $\kappa^2 = 32\pi G$. Then $\kappa=2/M$, where $M$ is the reduced Planck mass. Our conventions are $R_{\mu\nu}=R^\alpha_{\ \mu\alpha\nu}$, and $[\nabla_\mu,\nabla_\nu]v^\lambda = R_{\mu\nu\phantom{\lambda}\sigma}^{\phantom{\mu\nu}\lambda}v^\sigma$.}
\be 
\label{EH}
S_{EH} =  -2\int\!\! d^dx \, \sqrt{g} R/\kappa^2\,,
\ee
is unbounded from below,  the Euclidean signature partition function
\be 
\label{Z}
\mathcal{Z} = \int\!\! \mathcal{D}g_{\mu\nu}\ {\rm e}^{-S_{EH}}\,,
\ee
is not even na\"\i vely convergent \cite{Gibbons:1978ac}. If this embarrassment is to make any sense in an exact theory (as opposed to treating quantum gravity merely as an effective theory), it must be that this so-called ``conformal factor instability'' can somehow be reinterpreted. We will show that for particular parametrisations of the conformal factor and in a wide class of gauges, it indeed can be: it can be seen as supplying a UV  (ultraviolet) regularisation, rendering the theory finite in any dimension $d>2$. These properties extend to the full phenomenologically relevant action, thus including cosmological constant term and matter fields, provided that gauge fields are parametrised appropriately and massive tadpole corrections are set to zero by a trivial renormalisation (for example by normal ordering).

Let us emphasise that, apart from these simple provisos, we do not inject any new ingredients into the theory. The result just follows from performing partial resummations of the perturbation series in $\kappa$. However the resulting terms are non-perturbative in $\kappa$, leaving no obvious expansion parameter.
A related problem is that these terms fail individually to be invariant under BRST transformations for diffeomorphisms, or under background diffeomorphisms. We show that one can choose gauges in such a way that the gauge parameter itself controls the size of these terms, but not such that  higher orders terms are successively smaller or such that one can recover  diffeomorphism invariance term by term.

It is in fact an old speculation that quantum gravity should somehow provide a regularisation of the UV divergences one finds in quantum field theory \cite{Deser:2022taw,Deser:1957zz}, for example through the production of black holes \cite{Dvali:2014ila}. It has also been argued that the spinfoam realisation of Loop Quantum Gravity approach is naturally UV finite \cite{Perez:2000bf,Ashtekar:2004eh}. We do not see any substantive connection between these ideas and the results reported here. Since the formulation described here has no UV divergences, the renormalization group will play no r\^ole. Thus there is also no obvious connection to the asymptotic safety approach \cite{Weinberg:1980,Reuter:1996}. \TRM{We leave to the Conclusions (sec. \ref{sec:conclusions}), a discussion of earlier attempts \cite{DeWitt:1964yh,Isham:1972pf,tHooft:2010xlr} with closer connections to the ideas in this paper.}

\TRM{Comments on some approaches \cite{Einstein1919,Unruh1989a,Starobinsky:1980te} that are actually excluded in this framework are given at the end of sec. \ref{sec:two}. In that section} we set out the two key properties that we need, namely that all interaction terms are weighted by a positive exponential of the conformal factor, and that the latter propagates with the wrong sign. We show how this can be achieved for pure gravity, and for gravity coupled to matter fields, in all dimensions $d>2$. We note that fundamental higher derivative gravitational terms, such as the Starobinsky term \cite{Starobinsky:1980te}, are however excluded.  Then in sec. \ref{sec:partialresum} we demonstrate how the perturbation series in $\kappa$ can be resummed into UV finite terms, $\Gamma_n$, up to the trivial renormalisations mentioned earlier. In sec. \ref{sec:first}  we comment on how the latter ameliorate the so-called fine-tuning problem, while in sec. \ref{sec:second} we add some side remarks on why analytic continuation \cite{Gibbons:1978ac} does not alter the result, on an alternative point-splitting regularisation, and on more general parametrisations of the $\ph$ dependence.
In sec. \ref{sec:reorganise} we provide the proof of the compact formulae for the $\Gamma_n$, which we used in the previous section. In sec. \ref{sec:BRST} we demonstrate that with appropriate parametrisations, BRST invariance, more precisely the Zinn-Justin equations, can also be formulated as a UV finite expansion over the $\Gamma_n$. Then in sec. \ref{sec:curved} we consider quantisation around curved space. We show that it is straightforward to do so whilst preserving the UV finiteness of the corresponding $\Gamma_n$. However at this stage we encounter a problem: we show that background diffeomorphism invariance is not recovered by any finite resummation of the $\Gamma_n$. It is clear that the Zinn-Justin identities will similarly be violated at any finite order. As we discuss in sec. \ref{sec:conclusions}, this problem is potentially profound particularly because we no longer have an obvious parameter to control the expansion. By choosing the most general gauge at the bilinear level, we show in sec. \ref{sec:limits} that a particular gauge parameter $\xi$ controls the strength of the UV regularisation such that $\xi\to0$ ($\infty$) corresponds to making the regularisation infinitely weak (strong). However we see in detail that neither limit allows us to recover a controlled expansion. Finally in sec. \ref{sec:conclusions} we summarise and draw our conclusions, in particular we point to the existence of simpler UV finite models whose study might allow further progress.

\section{Two key properties}
\label{sec:two}

To linear order in $\kappa$, the metric $g_{\mu\nu}$, expanded around flat space, is given by
\be g_{\mu\nu} = \delta_{\mu\nu}+\kappa H_{\mu\nu} +O(\kappa^2)\,,\ee
where this defines the fluctuation field, $H_{\mu\nu}$.
We split out its trace, writing:
\be\label{trace} H_{\mu\nu} = h_{\mu\nu}+\tfrac2d \,\ph\,\delta_{\mu\nu}\,,\qquad\text{where}\quad \ph =\half \delta^{\mu\nu}H_{\mu\nu}\,,\quad\text{and}\quad \delta^{\mu\nu}h_{\mu\nu} =0\,.\ee
Non-perturbatively in $\kappa$, we parametrise the metric as
\be\label{paramg} g_{\mu\nu} = \mathrm{e}^{2\kappa \ph/d}\hat{g}_{\mu\nu}\,,\ee
where $\hat{g}_{\mu\nu}$ is \TRM{some fixed function of only the traceless part,} $h_{\mu\nu}$, and to be consistent with above must satisfy
\be\label{paramhg} \hat{g}_{\mu\nu} = \delta_{\mu\nu}+\kappa h_{\mu\nu} + O(\kappa^2)\,,\ee
but we leave it otherwise arbitrary. We note however that this allows for parametrisations of the metric which are non-singular for all real values of $\ph$ and $h_{\mu\nu}$, such as the exponential parametrisation considered \TRM{ in ref. \cite{Isham:1972pf}, see also \eg refs. \cite{Aida:1994zc,Nink:2014yya,Percacci:2015wwa}.}
Substituting parametrisation \eqref{paramg} into the Einstein-Hilbert action \eqref{EH}, gives
\be\label{EHreparam} S_{EH} = -2\int\!\! d^dx \, \sqrt{\hat{g}} \,\mathrm{e}^{\kappa(d-2)\ph/d}\left\{ \frac1{\kappa^2}\hat{R}+\frac{(d-1)(d-2)}{d^2}\hat{g}^{\mu\nu}\partial_\mu\ph\partial_\nu\ph\right\}\,.
\ee
Two key properties that we need are that the interaction terms are all weighted by a positive exponential of $\ph$, which we see is true here for all $d>2$, and that $\ph$ propagates with the wrong sign. 

This latter property is natural given that $\ph$ has the wrong sign kinetic term in \eqref{EHreparam}, but in fact whether it propagates with the wrong sign is a gauge dependent statement.\footnote{Indeed, recall that in Minkowski signature $\ph$ can be locally eliminated by a gauge choice. \TRM{The propagating asymptotic physical states are the usual ones, namely the transverse traceless $+$ and $\times$ polarisations.}} For example for gauge fixing, we can add to the action
\be\label{flatgf}\int\!\! d^dx \,\left\{ \frac{\alpha}{2}\delta^{\mu\nu}F_\mu F_\nu -\frac{1}{\kappa}\bar{c}^\mu QF_\mu\right\}\,, \ee
where 
\beal F_\mu &= \frac1\kappa \delta^{\alpha\beta} \left(\partial_\alpha g_{\beta\mu}-\frac12\partial_\mu g_{\alpha\beta} \right) = \delta^{\alpha\beta}\partial_\alpha h_{\beta\mu}-\left(\frac2d-1\right)\partial_\mu\ph +O(\kappa)\nn\\
&= \delta^{\alpha\beta} \mathrm{e}^{2\kappa \ph/d} \left(\frac1\kappa\partial_\alpha \hat{g}_{\beta\mu}-\frac1{2\kappa}\partial_\mu \hat{g}_{\alpha\beta} +\frac2d\partial_\alpha\ph\,\hat{g}_{\beta\mu}-\frac1d\partial_\mu\ph\,\hat{g}_{\alpha\beta}\right)\,,\label{F}
\eeal
is DeDonder gauge (plus interactions when written in terms of the fluctuation fields)
and $\alpha$ is a gauge parameter. The second term in \eqref{flatgf} is the ghost action,  where
\be 
\label{Qg}
Qg_{\mu\nu} = 2\kappa \partial_{(\mu} c^\alpha g_{\nu)\alpha} +\kappa c^\alpha \partial_\alpha g_{\mu\nu} = \kappa\,\mathrm{e}^{2\kappa \ph/d} \left(
2\partial_{(\mu} c^\alpha \hat{g}_{\nu)\alpha} +2\frac{\kappa}{d}c^\alpha\partial_\alpha\ph\,\hat{g}_{\mu\nu}+c^\alpha\partial_\alpha\hat{g}_{\mu\nu}\right)
\ee
is the BRST transformation of the metric, which is just its Lie derivative along $\kappa c^\mu$. Here, $c^\mu$ and $\bar{c}^\mu$ are the ghost and antighost respectively. The combined action \eqref{EHreparam} and \eqref{flatgf} has the usual BRST invariance, the remaining BRST transformations being
\be \label{Qghost} Qc^\mu = \kappa c^\nu\partial_\nu c^\mu\,,\qquad Q\bar{c}^\mu = \alpha\, \delta^{\mu\nu} F_\nu\,.\ee
(As usual $Q$ is nilpotent but only on shell, because $Q^2\bar{c}^\mu = \alpha\, \delta^{\mu\nu} QF_\nu$ vanishes only after using the $\bar{c}$ equations of motion. We could have also introduced an auxiliary field $b^\mu$ for full off-shell nilpotency, following \eg ref. \cite{Mandric:2023dmx}.)

\TRM{At first sight the parametrisation \eqref{paramg} is dangerously ambiguous, introducing a new gauge transformation where a change $\delta\ph$ can be compensated by a change $\delta h_{\mu\nu}$. This would remain unfixed by \eqref{F}, so one would have to seek to eliminate it by imposing for example that $\hat{g}_{\mu\nu}$ has unit determinant. Actually the restriction \eqref{paramhg} eliminates such a gauge transformation to any order in $\kappa$. To see this note that, from expanding (2.3), a small shift $\delta\varphi$ would have to be compensated by a small change $\delta h_{\mu\nu}$ such that:
\be 0 = \frac2d\,\delta\varphi\, \delta_{\mu\nu} + \delta h_{\mu\nu} +[O(\kappa)(\delta h\ \mathrm{or}\ \delta\varphi)]_{\mu\nu}\,.\ee
Then the lowest order of $\kappa$ at which $\frac2d\,\delta\varphi\, \delta_{\mu\nu}$ is non-vanishing, must be cancelled by a part of the same order in $\delta h_{\mu\nu}$. But this is impossible because $\delta h_{\mu\nu}$ is traceless. A related point is that the gauge fixing we have done is sufficient to allow the kinetic terms to be inverted and give the propagators.}

\TRM{In} momentum space the propagators for $h_{\mu\nu}$ and $\ph$ take the form (see \eg \cite{Morris:2018axr}):
\beal
\label{ppalph}
\langle \ph(p)\,\ph(-p)\rangle &= \left(\frac1\alpha - \frac{d-1}{d-2}\right) \frac1{p^2}\,,
\\
\label{hpalph}
\langle h_{\mu\nu}(p)\, \ph(-p)\rangle &= \langle \ph(p)\,h_{\mu\nu}(-p)\rangle =
\left(1-\frac2\alpha\right) \left( \frac{\delta_{\mu\nu}}{d}-\frac{p_\mu p_\nu}{p^2}\right) \frac1{p^2}\,,\\
\langle h_{\mu\nu}(p)\,h_{\alpha\beta}(-p)\rangle &= \frac{\delta_{\mu(\alpha}\delta_{\beta)\nu}}{p^2} 
+\left(\frac4\alpha-2\right)\frac{p_{(\mu}\delta_{\nu)(\alpha}p_{\beta)}}{p^4}
+\frac1{d^2} \left(\frac4\alpha-d-2\right) \frac{\delta_{\mu\nu}\delta_{\alpha\beta}}{p^2}\nn\\
\label{hhalph}
&\qquad\qquad\qquad\qquad+\frac2d\left(1-\frac2\alpha\right)\frac{\delta_{\alpha\beta}p_\mu p_\nu+p_\alpha p_\beta \delta_{\mu\nu}}{p^4}\,.
\eeal
Thus to ensure we have the key property that $\ph$ has a wrong-sign propagator, we just need to choose $\alpha$ outside the range $0<\alpha< (d-2)/(d-1)$. Note that for all dimensions $d>3$ this includes the popular Feynman -- DeDonder gauge, $\alpha=2$, where $h_{\mu\nu}$ and $\ph$ decouple and the propagators simplify significantly. In fact in this case their kinetic terms are simply
\be\label{kinetic} \frac12(\partial_\mu h_{\alpha\beta})^2-\frac{d-2}{d}(\partial_\mu\ph)^2\,.\ee
In $d=4$ dimensions both these kinetic terms are canonically normalised; this is why we defined $\ph$ with the factor of a half in \eqref{trace}.

Notice that the gauge fixing term interactions and ghost interactions in \eqref{flatgf}, are also weighted by positive exponentials of $\ph$, thanks to its appearance in \eqref{F} and \eqref{Qg}.
So far we have only considered the quantisation of the Einstein-Hilbert action, but now let us show that we can ensure that all interactions in a full phenomenologically relevant theory of quantum gravity, are weighted by a positive exponential of $\ph$. First note that we can add a cosmological constant term since 
\be \label{cc}\sqrt{g} = \mathrm{e}^{\kappa \ph} \sqrt{\hat{g}}\ee
is weighted by a positive exponential of $\ph$.  Thanks to the $\sqrt{g}$ factor, scalar field mass and interaction terms will have the same exponential, whilst their kinetic terms pick up the same exponential as in \eqref{EHreparam} due to the presence of $g^{\mu\nu}$ in the contraction over derivatives of the scalar field. \TRM{The mass term for  fermions, carries the \eqref{cc} exponential, as does a Yukawa interaction with scalars. The kinetic term for fermions is also weighted by a positive exponential:
\be\label{fermion} \sqrt{g}\,\bar{\psi}(i\gamma^\mu\nabla_\mu+m)\psi = \mathrm{e}^{\kappa (d-1)\ph/d} \sqrt{\hat{g}}\,\bar{\psi}i\hat{\gamma}^\mu\left(\hat{\nabla}_\mu +\frac{\kappa}{2d}(d-1)\partial_\mu\ph\right)\psi +\mathrm{e}^{\kappa \ph} \sqrt{\hat{g}}\,\bar{\psi}m\psi\,.\ee
To see this, note that the spin connection in $\nabla_\mu$ does not carry an overall exponential. This follows because each term in the spin connection contains as many instances of the vielbein $e^a_\mu=\mathrm{e}^{\kappa \ph/d}\, \hat{e}^a_\mu$ as its inverse $e^\mu_a = \mathrm{e}^{-\kappa \ph/d}\,\hat{e}^\mu_a$. For a similar reason the Levi-Civita connection also does not carry an overall exponential. However as written, the $\gamma$ matrix contains the inverse veilbein: $\gamma^\mu = \gamma^a e^\mu_a$. Substituting these explicit formulae and those for the connection, then gives the right hand side of \eqref{fermion}.}

Vector fields $A_\mu$ are apparently a problem, not in terms of interactions with scalars or fermions, but because this requires adding for their kinetic terms 
\be\label{A} \frac14\sqrt{g}\, g^{\mu\sigma} g^{\nu\lambda} F_{\mu\nu} F_{\sigma\lambda} = \frac14\sqrt{\hat{g}}\,\mathrm{e}^{\kappa(d-4)\ph/d}\, \hat{g}^{\mu\sigma} \hat{g}^{\nu\lambda} F_{\mu\nu} F_{\sigma\lambda}\,,\ee
where $F_{\mu\nu}$ is the corresponding field strength. In this case the positive $\ph$ exponential would be missing in the phenomenologically relevant case of $d=4$ dimensions. However we can easily repair this by writing $A^\mu$ as the fundamental field, so that in the field strength we now have $A_\mu = g_{\mu\nu} A^\nu$. Then the kinetic term in \eqref{A} is weighted by $\mathrm{e}^{\kappa \ph}$, whilst the cubic and quartic $A$ interactions (in the non-Abelian case) have even more positive exponentials. Note also that this change of variables only improves the situation for interactions with scalars or fermions. Another fix is to define $A_\mu$ to transform as a density of weight \TRM{$w>(d-4)/d$}, then what would appear in the action is $g^{w/2}A_\mu$. This covers all the fields needed for the Standard Model and its extensions.\footnote{\TRM{However one also has to gauge fix such gauge field symmetries, and one may wonder if that introduces any difficulties. It does not, as we show in app. \ref{app:YM}.}}

At first sight, higher derivative terms are excluded in $d=4$ dimensions, for example
\be \sqrt{g}\left(g^{\mu\nu}\partial_\mu s \partial_\nu s\right)^n\,,\ee
where $n\ge2$ and $s$ is a scalar field, since they will end up with the weighting $\mathrm{e}^{\kappa(d-2n) \ph/d}$. That could be regarded as a welcome outcome since higher derivative terms are often problematic anyhow, for reasons of stability and unitarity. However we can also fix these up with a positive exponential of $\ph$ by defining $s$ to be a density with a suitably positive weight. 

In $d=4$ dimensions, purely gravitational higher derivative terms are however genuinely excluded. In particular $\sqrt{g}R^2$ would end up with weighting $\mathrm{e}^{\kappa(d-4) \ph/d}$. This is the so-called Starobinsky term, a physically acceptable modification of Einstein's gravity \cite{Starobinsky:1980te} which describes inflation and provides one of the best matches to the Planck data \cite{Planck:2018jri}.  Although we need to exclude the Starobinsky term, inflation can still be described through the equivalent scalar potential in the Einstein frame, or indeed through any preferred choice of inflaton potential, since these all have the same weighting as the cosmological constant term \eqref{cc}.

Note that changing integration variable in the pure gravity sector has no effect on the conclusions above. For example one might try to quantise with $g^{\mu\nu}$ or a densitised version for example $\mathfrak{g}^{\mu\nu}=\sqrt{g} g^{\mu\nu}$, parametrising these with an appropriate $\ph$ exponential multiplied by a purely $h_{\mu\nu}$ dependent part. However the coefficient in the $\ph$ exponential is fixed by normalisation of its kinetic term, \eg as in \eqref{kinetic}, and thus re-expressing the metric $g_{\mu\nu}$ in terms of these new variables just leads us back to our parametrisation \eqref{paramg} again. On the other hand, some alternative approaches to quantum gravity are evidently excluded in this scheme, including unimodular gravity \cite{Einstein1919,Unruh1989a}, the latter because it does not have a conformal factor.

\section{Partial resummation into UV finite terms}
\label{sec:partialresum}

The partial resummation of perturbation theory into a series of UV finite terms, $\Gamma_n$, is achieved as follows. We split off from the action all the kinetic terms (the terms bilinear in the fields) which will give us the propagators, but crucially we do not expand the exponentials in $\ph$. For example for the Einstein-Hilbert term \eqref{EHreparam}, this will give us the following interaction vertex
\besp\label{EHvertex} -2 \sqrt{\hat{g}} \,\mathrm{e}^{\kappa(d-2)\ph/d}\left\{ \frac1{\kappa^2}\hat{R}+\frac{(d-1)(d-2)}{d^2}\hat{g}^{\mu\nu}\partial_\mu\ph\partial_\nu\ph\right\}\\ -\frac12(\partial_\mu h_{\alpha\beta})^2+(\partial^\mu h_{\mu\nu})^2+\frac2d(d-2)\ph\partial^2_{\mu\nu}h^{\mu\nu}+2\frac{(d-1)(d-2)}{d^2}(\partial_\mu\ph)^2\,.\eesp
If desired, this vertex can be broken into smaller pieces, provided that in each piece the exponential is not expanded and its corresponding bilinear terms are subtracted. 

We then perform perturbation theory on these vertices, again without expanding the exponentials, and resum all the quantum corrections involving only $\ph$ propagators. We will see that the resulting terms, $\Gamma_n$, are all UV finite, and furthermore this finiteness property is preserved to all orders in perturbation theory for the remaining quantum corrections, in pure quantum gravity being those involving $h_{\mu\nu}$ or the ghosts which we treat perturbatively in $\kappa$. In this section and for the most part of the paper, these remaining quantum corrections will be taken also to include those containing the mixed propagators,  $\langle\ph\, h_{\mu\nu}\rangle$. However in sec. \ref{sec:limits}, we consider a limit where one is led to resum also these corrections.

If (beyond) the standard model matter fields are included, they are treated similarly. However for these one has a natural way to split the action into smaller pieces since matter-matter interactions are proportional to their own couplings which can be treated perturbatively. Again, we do so without expanding the $\ph$ exponentials however.

We proceed rigorously as follows. We split off from the classical action all the kinetic terms $\half\phi^A\Delta^{-1}_{AB}\phi^B$, treating the remaining interaction part of the classical action $S_I[\TRM{\phi}]$ as a single unit. Here we are using compact DeWitt notation, so Einstein summation over the capital indices indicates both summation over Lorentz indices and integration over spacetime. The field $\phi^A$ stands for all the fields we have to integrate over, both fermionic and bosonic. We treat $S_I$ as a single unit by introducing an overall vertex counting parameter $\epsilon$, setting $S_I[\phi]\mapsto \epsilon S_I[\phi]$. We then expand perturbatively in $\epsilon$, acknowledging at the end of the process that actually $\epsilon=1$. For the Legendre effective action
\be\label{Leg} \Gamma[\Phi] = \half \Phi^A\Delta^{-1}_{AB}\Phi^B+\Gamma_I[\Phi]\ee 
($\Phi^A$ are the corresponding so-called classical fields), the resulting series takes the form
\be\label{epsexp} \Gamma_I[\Phi] = \sum_{n=1}^\infty \epsilon^n \Gamma_n\,. \ee
In sec. \ref{sec:reorganise}, we derive expressions for the $\Gamma_n$, \TRM{using functional analysis. Since sec. \ref{sec:reorganise} stands on its own, it can if preferred be read now.}
Here we just quote the results, since the combinatorics involved are readily understood. We will see that the $n^\text{th}$ order correction can be written in terms of $n$ copies of $\Gamma_1$ multiplied together, \viz $\Gamma_1^n$, and a differential operator $\Po_{ij}$ that acts on individual copies in this power:
\be \label{Poij} \Po_{ij} = \Delta^{AB}\frac{\partial}{\partial\Phi^B_i}\frac{\partial}{\partial\Phi^A_j}\,. \ee
Here the first differential acts only the $i^\text{th}$ copy, whilst the second differential acts on the $j^\text{th}$ copy.

\subsection{First order}
\label{sec:first}

\begin{figure}[ht]
\centering
\includegraphics[scale=0.35]{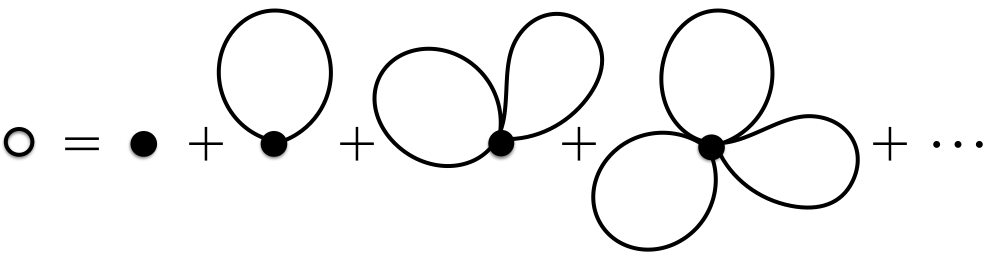}
\caption{The classical interaction vertex is represented by the black dot and has any number of external legs (which are not drawn). Added to this is all its 1PI quantum corrections \ie the sum over all tadpole corrections. }
\label{fig:tadpoles}
\end{figure}

To first order, \viz $\Gamma_1$, we keep only one copy of $S_I$. For the Legendre effective action the sum over all 1PI (one-particle irreducible) quantum corrections thus takes the form of a tadpole expansion as illustrated in fig. \ref{fig:tadpoles}. Using \eqref{Poij}, it can be written compactly as \TRM{(see sec. \ref{sec:reofirst})}:
\be\label{Gone} \Gamma_1 =  \mathrm{e}^{\Po_{11}/2} S_I[\Phi] \,,\ee
The next step is to perform the sum over the $\ph$ quantum corrections. For a gauge invariant regulator such as dimensional regularisation, this is trivial since massless tadpole integrals are set to zero. For pure quantum gravity, the remaining quantum corrections are again massless tadpole integrals, involving $h_{\mu\nu}$ and the ghosts, and thus they also vanish. Thus in fact in this case the operator $\mathrm{e}^{\Po_{11}/2}$ has no effect and $\Gamma_1 = S_I$.   Other regulators require care in order \TRM{to} recover gauge invariance, but for example one can get the same answer in a much older framework by normal ordering the $S_I$ vertices since then the vacuum expectation value of the vertex automatically vanishes. 

On the other hand, if massive matter fields are included, their tadpoles do not vanish and furthermore they are UV divergent. This is the one type of correction where no UV regularisation is supplied dynamically by the conformal factor. However such tadpole corrections are trivial in that they are local and just result in (divergent) redefinitions of the higher mass-dimension couplings in the theory. For example for a massive scalar field $\phi$, the $\phi$-tadpole correction involving the quartic interaction $\sqrt{g}\lambda\phi^4$ leads to a divergent correction to its mass term $\half\sqrt{g} m^2\phi^2$ in the form, $\delta m^2\propto \lambda m^2$. These divergent corrections to higher mass-dimension couplings, can be removed either by a trivial renormalisation or, as above, by normal ordering. In practice then, we can ignore them.

In the light of the phenomenological ``fine-tuning" problems of the Higgs' mass and cosmological constant, it is intruiging that their massive tadpole corrections, which form by far the largest contribution, must be trivially removed in this way for consistency of this theory. We note however that it is still the case that the Higgs' mass and cosmological constant parameters in the action $S$ would differ by finite corrections from the corresponding measurable quantities. Furthermore, these corrections can be expected to be large since, as we see explicitly in sec. \ref{sec:limits}, they involve quantum corrections cut off in the UV by the Planck mass. \TRM{In any case, the real questions to answer are dynamical ones, for the cosmological constant for example relating its IR value at two different times -- namely during inflation and now, and answering such questions would require a more complete theory.}

\subsection{Second order}
\label{sec:second}

\begin{figure}[ht]
\centering
\includegraphics[scale=0.4]{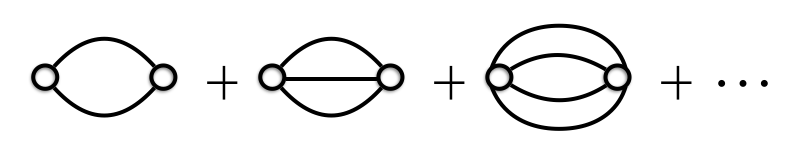}
\caption{The second-order part, $\Gamma_2$, of the Legendre effective action, is given by an expansion over melonic Feynman diagrams, where the open circles are copies of $\Gamma_1$ as given in fig. \ref{fig:tadpoles}. Again, these copies have any number of external legs which are not drawn.}
\label{fig:melons}
\end{figure}

The second order correction has the expansion illustrated in fig. \ref{fig:melons}. It is given by
\be\label{melons} \Gamma_2[\Phi] = -\frac12 \left(\mathrm{e}^{\Po_{12}}-1-\Po_{12}\right) \Gamma_1^{\,2}\,.\ee
\TRM{It is proven in sec. \ref{sec:reosecond}, but the} combinatorics in this result can be readily understood. In particular the factor of $\half$ comes from expanding the exponential of $S_I$ to second order. Then propagators are distributed in all possible ways. This turns the instances of $S_I$ into instances of the tadpole expansion \eqref{Gone}, \ie $\Gamma_1$, and these two instances are connected by propagators, giving the melonic expansion illustrated in fig. \ref{fig:melons}. The $-1-\Po_{12}$ correction in the above simply serves to remove the 1PR (one-particle reducible) and disconnected diagrams.

Since the $-1$ and $-\Po_{12}$ correction terms are disconnected and tree-level respectively, they clearly do not contribute UV divergences. Meanwhile, $\Gamma_1$ takes the form
\be\label{Goneform} \Gamma_1 =\sum_\alpha \int\!\!d^dx \,\mathrm{e}^{\kappa\beta_\alpha \ph} \mathcal{L}_\alpha(\ph,\partial\ph)\,,\ee
plus explicit bilinear terms. However the latter contribute only to the first diagram in fig. \ref{fig:melons} and, for pure quantum gravity, collapse it to a massless tadpole integral which again vanishes in dimensional regularisation. For the massive matter fields, the result is a UV divergent tadpole integral which again we can discard by a trivial renormalisation.
Thus to show that the non-trivial part is UV finite, we need concentrate only on the terms in \eqref{Goneform} and on the exponential of $\Po_{12}$ in \eqref{melons}.

In \eqref{Goneform} the sum runs over the different $\beta_\alpha>0$. In an abuse of notation, $\ph\in \Phi^A$ now stands for its classical field counterpart. We are suppressing the dependence of the Lagrangian densities $\mathcal{L}_\alpha$ on all the other fields. For simplicity we have written that the $\mathcal{L}_\alpha$ depend at most on single derivatives $\partial_\mu\ph$. This can be arranged but is not essential. The generalisation to higher derivatives is straightforward.

The next step is to sum up all the purely $\ph$ quantum corrections in \eqref{melons}. For this it is helpful to work in position space. The canonical massless scalar propagator is
\be\label{prop} \Delta(x) = \int\!\!\frac{d^dp}{(2\pi)^d} \frac{\mathrm{e}^{ip\cdot x}}{p^2} = \frac{\Gamma(d/2)}{2\pi^{d/2}(d-2)}\frac{1}{x^{d-2}}\,.
\ee
We are working in a gauge such that  the $\ph$ propagator is $-\xi^2\Delta(x)$, for some $\xi$ which is real and non-vanishing.\footnote{\label{foot:xi}For example, from \eqref{ppalph} it is given by $\xi= \sqrt{(d-1)/(d-2)-1/\alpha}$, while in Feynman gauge this becomes $\xi=\sqrt{d/2(d-2)}$.} 
Inspired by standard tricks with the functional integral, we can rewrite \eqref{Goneform} in a more convenient way:
\be\label{GoneformJ} \Gamma_1 =\sum_\alpha \int\!\!d^dx \, \mathcal{L}_\alpha\left(\frac{\partial}{\partial J_x},\partial\frac{\partial}{\partial J_x}\right)\, \mathrm{e}^{(\kappa\beta_\alpha+J_x) \ph(x)}\Big|_{J_x=0}\,.\ee
Here $J_x$ carries $x$ dependence but these are genuinely partial derivatives so that 
\be\frac{\partial}{\partial J_x}J_x=1\,.\ee
Now, substituting \eqref{GoneformJ} into \eqref{melons}, the sum over purely $\ph$ corrections is straightforward to evaluate:
\besp\label{Gtworesum} \Gamma_2[\Phi] = -\frac12\,\mathrm{e}^{\hat{\Po}_{12}}\sum_{\alpha_1,\alpha_2}\int\!\!d^dx_1d^dx_2\,\mathcal{L}_{\alpha_1}\left(\frac{\partial}{\partial J_{x_1}},\partial\frac{\partial}{\partial J_{x_1}}\right)\mathcal{L}_{\alpha_2}\left(\frac{\partial}{\partial J_{x_2}},\partial\frac{\partial}{\partial J_{x_2}}\right)\\
\exp\big\{[\kappa\beta_{\alpha_1}+J_{x_1}]\ph(x_1)+[\kappa\beta_{\alpha_2}+J_{x_2}]\ph(x_2)-\xi^2[\kappa\beta_{\alpha_1}+J_{x_1}]\Delta(x_1-x_2)[\kappa\beta_{\alpha_2}+J_{x_2}]\big\}\Big|_{J=0}\,,
 \eesp
where $\hat{P}_{ij}$ is the operator \eqref{Poij} with the purely $\ph$ piece ($A=B=\ph$) removed. The important observation is that all corrections end up  weighted by the exponential
\be \label{regulate}
\exp\big\{-\xi^2\kappa^2\beta_{\alpha_1}\beta_{\alpha_2}\Delta(x_1-x_2)\big\}
\ee
which results from resumming all the pure $\ph$ quantum corrections.
We still have some $J$ differentials to perform in the above before setting $J=0$, which will bring down $\Delta(x_1-x_2)$ factors, and any number of further propagator factors appear for the other field combinations as encoded in $\hat{P}_{12}$ and which we treat perturbatively. 
In position space UV divergences arise from the fact that as $x_1\to x_2$, the propagator $\Delta(x_1-x_2)$ diverges as a power, \viz as $1/|x_1-x_2|^{d-2}$. Larger powers appear on applying the spacetime derivatives in $\mathcal{L}_\alpha$. But, thanks to the overall minus sign in the exponential in \eqref{regulate}, the resummed pure $\ph$ quantum corrections supply an exponential damping which is strong enough to overcome any power-law divergence in the multiplying factors. Thus we see that all these quantum corrections actually vanish exponentially fast as $x_1\to x_2$.
We have therefore established that the wrong sign propagator for $\ph$ combined with the positive exponentials of $\ph$ in every interaction, provide a non-perturbative dynamical mechanism that regulates all remaining quantum corrections in $\Gamma_2$.

As a concrete example we take the Einstein-Hilbert vertex \eqref{EHvertex} and compute the remaining $J$ differentials in \eqref{Gtworesum} to explicitly the complete resummation over purely $\ph$ quantum corrections:
\beal \Gamma_2\ \ni \ &-2\,\mathrm{e}^{\hat{\Po}_{xy}}\int\!\!d^dx d^dy\, \sqrt{\hat{g}(x)}\sqrt{\hat{g}(y)}\,\mathrm{e}^{\kappa\beta[\ph(x)+\ph(y)]}\,\mathrm{e}^{-\xi^2\beta^2\kappa^2\Delta}\,\Big\{ \nn\\
&\frac1{\kappa^4}\hat{R}(x)\hat{R}(y)+2\frac{\sigma}{\kappa^2}\hat{R}(x)\hat{g}^{\alpha\beta}(y)\partial^y_\alpha\left[\ph(y)-\xi^2\beta\kappa\Delta\right]
\partial^y_\beta\left[\ph(y)-\xi^2\beta\kappa\Delta\right]\nn\\
&+\sigma^2\hat{g}^{\mu\nu}(x)\,\hat{g}^{\alpha\beta}(y)\,\partial^x_\mu\left[\ph(x)-\xi^2\beta\kappa\Delta\right]\partial^x_\nu\left[\ph(x)-\xi^2\beta\kappa\Delta\right]\partial^y_\alpha\left[\ph(y)-\xi^2\beta\kappa\Delta\right]
\partial^y_\beta\left[\ph(y)-\xi^2\beta\kappa\Delta\right]\nn\\
&+4\xi^2\sigma^2\hat{g}^{\mu\nu}(x)\,\hat{g}^{\alpha\beta}(y)\,\partial^x_\mu\partial^x_\alpha\Delta\,
\partial^x_\mu\left[\ph(x)-\xi^2\beta\kappa\Delta\right]\partial^y_\beta\left[\ph(y)-\xi^2\beta\kappa\Delta\right]\nn\\
&+2\xi^4\sigma^2\hat{g}^{\mu\nu}(x)\,\hat{g}^{\alpha\beta}(y)\,\partial^x_\mu\partial^x_\alpha\Delta\,\,\partial^x_\nu\partial^x_\beta\Delta\Big\}\,,\label{Gtwoexample}
\eeal
where $\Delta=\Delta(x-y)$, $\beta = (d-2)/d$, and for brevity we have also introduced $\sigma=(d-1)(d-2)/d^2$.
It is tedious but straightforward to confirm that the same answer is arrived at by computing explicitly the Feynman diagrams in fig. \ref{fig:melons} and then summing over them. We see again that, despite the products of (differentials of) propagators, as appear inside the braces above, all contributions are UV finite, thanks to the presence of $\mathrm{e}^{-\xi^2\beta^2\kappa^2\Delta}$. 

Treating perturbatively the remaining quantum corrections involving other fields, for \eqref{Gtwoexample} those that involve $h_{\mu\nu}$, will result in new terms with further factors of (differentials of) propagators, but does not alter the regularising exponential. Therefore all these (in fact infinitely many) further corrections are also UV finite. 

For matter fields we would want to split these contributions up perturbatively according to the power of their couplings (namely powers 0, 1, or 2, since we only have two copies of $S_I$), but since this does not harm the $\ph$ exponentials, these  pieces are evidently also UV finite.

Before considering third order, we make some further comments. 
In order to solve the problem of the lack of convergence of the functional integral \eqref{Z}, the authors of   ref. \cite{Gibbons:1978ac} advocate analytically continuing $\ph\mapsto i\ph$, so that the $\ph$ functional integral is performed for imaginary amplitudes. One can choose to do so here since the change variables does not affect this dynamical UV regularisation. In this case the $\ph$ propagator is positive, but the coefficient in the $\ph$ exponentials is now imaginary, \ie we have $\beta\mapsto i\beta$. Thus again we find that the quantum corrections are regulated by the factor \eqref{regulate}, the crucial minus sign in the exponential now arising from $(i\beta)^2$.

For pure gravity, we deduced that the sum over tadpoles at first order, \eqref{Gone}, is $\Gamma_1=S_I$, because all the tadpoles vanish in dimensional regularisation. However from the above analysis, in particular \eqref{regulate}, we see that the dynamical regularisation results in a prefactor of the form $\mathrm{e}^{-\half\xi^2\beta^2\kappa^2\Delta(0)}$. In a point splitting regularisation this prefactor would force $\Gamma_1$ and all higher order corrections to vanish, leaving us with just the bilinear terms in \eqref{Leg}, \ie a free theory. This result is incompatible with the BRST invariance (\ref{Qg},\ref{Qghost}), but point-splitting is not a gauge invariant regularisation. We can compensate by making an infinite renormalisation to cancel this prefactor, which would then take us back to the dimensional regularisation result $\Gamma_1=S_I$. On the other hand higher orders, for example \eqref{Gtwoexample}, are not affected by point-splitting since the result is already finite.

We parametrised the metric \eqref{paramg}, using an exponential of $\ph$, which in turn results in vertices having an exponential of $\ph$ as a factor. The exponential is unique in that vertices continue to have this property no matter how many times they are differentiated in the process of forming the corrections, $\Gamma_n$. It is possible to use other parametrisations for $\ph$ and resum all the propagator corrections, \eg the tadpoles and melonic contributions, see \eg refs. \cite{Morris:2018mhd,Morris:2020blt}. However the functional form of the $\ph$ dependence will then change depending on the order, $\Gamma_n$, and the results can be singular in IR or UV limits \cite{Morris:2018mhd,Morris:2020blt}.

\subsection{Third order}

\begin{figure}[ht]
\centering
\includegraphics[scale=0.4]{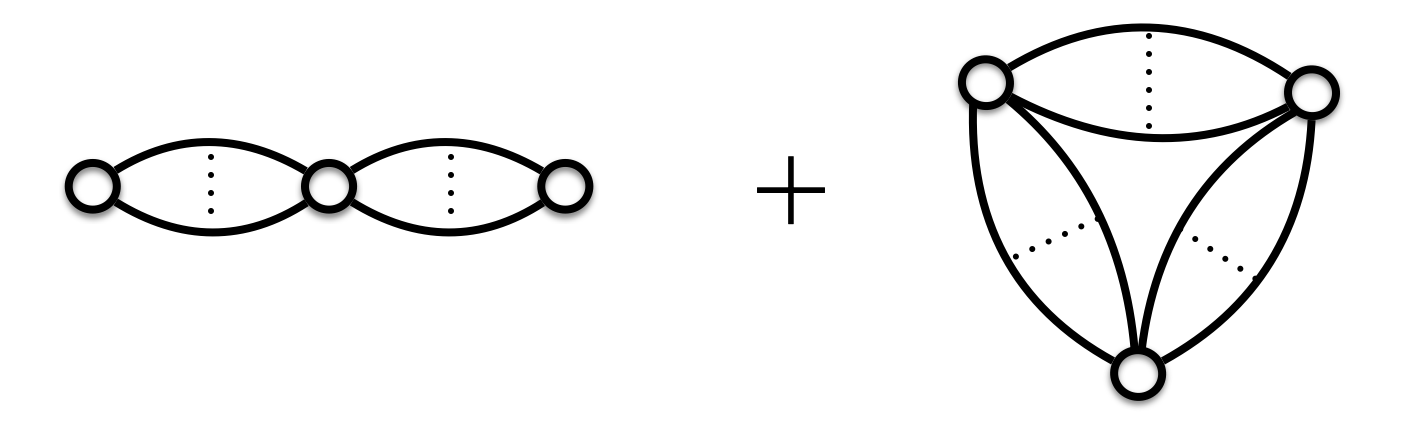}
\caption{The third-order part, $\Gamma_3$, joins three copies of $\Gamma_1$ together using melonic expansions. In the linear topology, each melonic expansion starts with two propagators as in fig. \ref{fig:melons}, thus excluding 1PR and disconnected diagrams. 
In the triangle topology, the melonic expansions start with a single propagator, and in this way include the remaining 1PI diagrams whilst also excluding diagrams that are 1PR or disconnected or of the linear topology.}
\label{fig:thirdorder}
\end{figure}

Now we consider the correction that is third order in $\epsilon$. It is given by 
\be\label{thirdorder} \Gamma_3 =\frac{1}{2!}\left(\mathrm{e}^{\Po_{12}}-1-\Po_{12}\right) \left(\mathrm{e}^{\Po_{23}}-1-\Po_{23}\right) \Gamma_1^{\,3}
+\frac{1}{3!}\left(\mathrm{e}^{\Po_{12}}-1\right)\left(\mathrm{e}^{\Po_{23}}-1\right)\left(\mathrm{e}^{\Po_{31}}-1\right)\Gamma_1^{\,3} \,,
\ee
and illustrated diagrammatically in fig. \ref{fig:thirdorder}. Again this formula is derived in sec. \ref{sec:reorganise} \TRM{(sec. \ref{sec:reothird})} but the combinatorics can be readily appreciated using the usual Feynman diagram systematics. 

In order to prove that the third order correction is UV finite once we sum over purely $\ph$ quantum corrections, we can again ignore the explicit  disconnected and 1PR pieces in \eqref{thirdorder}. To see this note that in the first contribution \viz the linear topology, if we retain just the $-1$ in one bracket, it disconnects, leaving just the melonic expansion of $\Gamma_2$ as in fig. \ref{fig:melons} that we have already shown is UV finite. Likewise if we retain just say the $-\Po_{12}$ term in the first bracket, it becomes 1PR with its loops regularised as in $\Gamma_2$. Turning now to the second contribution in \eqref{thirdorder}, \ie the triangle topology in fig. \ref{fig:thirdorder}, if we retain just a $-1$ from one of these brackets, the triangle disconnects into something that has the same UV properties as the linear topology.

Now, potential UV divergences arise in $\Gamma_3$ from the limit where any two $x$-integrands in the three $\Gamma_1$, are brought close to each other, or when all three meet at a single point. However, writing the copies of $\Gamma_1$ in the form \eqref{GoneformJ}, it is clear that the second order result \eqref{Gtworesum} generalises in the sense that each melonic expansion over purely $\ph$ corrections, provides the UV regulating term
\be \label{regulategen}
\exp\big\{-\xi^2\beta_{\alpha_i}\beta_{\alpha_j}\kappa^2\Delta(x_i-x_j)\big\}\,,
\ee
which ensures that all the quantum corrections are UV finite in these limits, in the same way as before. 

\begin{figure}[ht]
\centering
\includegraphics[scale=0.4]{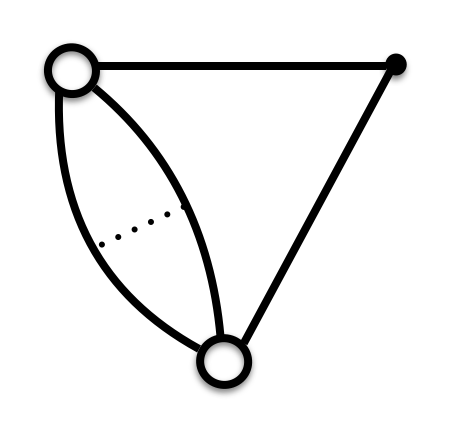}
\caption{Retaining the explicit bilinear terms in just one copy of $\Gamma_1$ of the triangle topology, as represented here by the dot, results in a non-vanishing contribution which however is UV regulated by the remaining melonic expansion.}
\label{fig:bilinear}
\end{figure}

As for the explicit bilinear terms in $\Gamma_1$, they play no r\^ole in the linear topology of fig. \ref{fig:thirdorder} for much the same reason as before. They do not contribute in the inner $\Gamma_1$ since it needs at least four legs, whilst if we retain just the bilinear terms in an outer $\Gamma_1$, it collapses to a tadpole that either vanishes or is removed by a trivial renormalisation.
Likewise if we retain only the bilinear terms in two of the $\Gamma_1$ copies in the triangle topology of fig. \ref{fig:thirdorder}, it collapses to a  tadpole that we treat in the same way. However if we retain the bilinear term in just one of the $\Gamma_1$ copies we see that the result is non-vanishing, as illustrated in fig. \ref{fig:bilinear}. But now the only UV divergences in this contribution can come from the integrands in the remaining pair of $\Gamma_1$ coming close to each other, and this limit is still regulated by its melonic sum factor \eqref{regulategen}.

\begin{figure}[ht]
\centering
\includegraphics[scale=0.4]{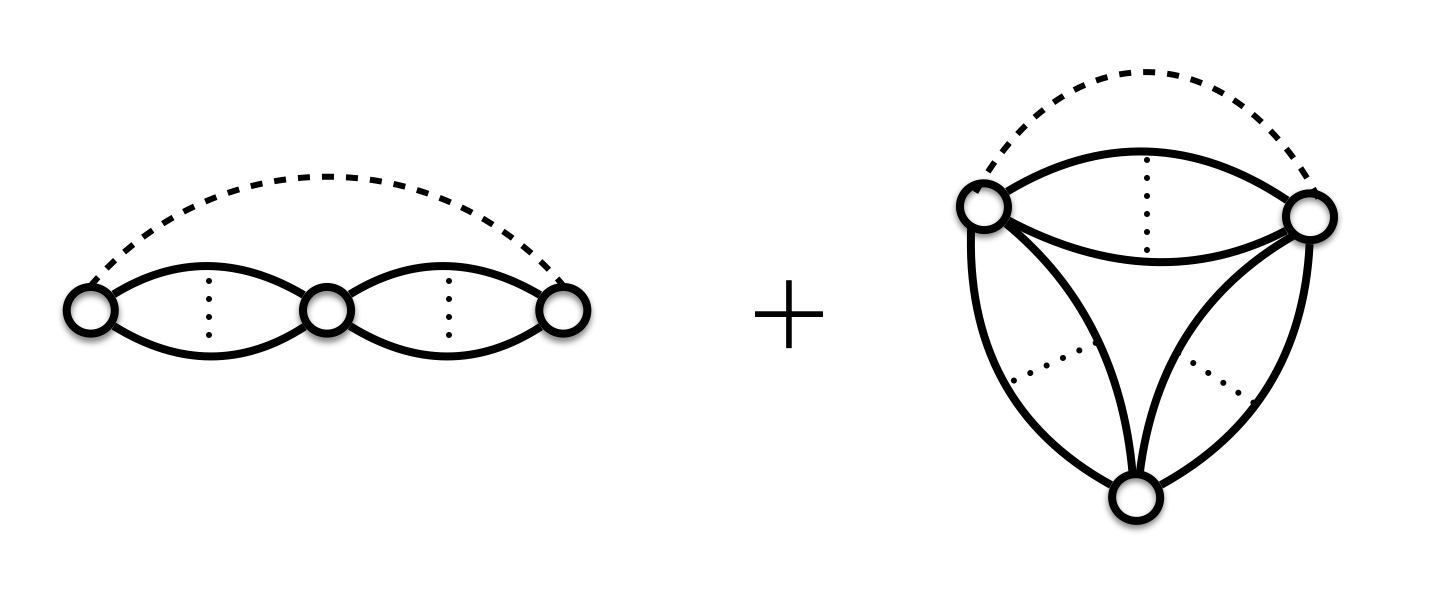}
\caption{We treat perturbatively the further quantum corrections, \ie those that involve $h_{\mu\nu}$ and/or ghosts. Those corrections involving propagators attaching the outer copies of $\Gamma_1$ in the linear topology, as in the first figure, appear at first sight not to be UV regularised by a melonic expansion. However the combinatorics are such that these quantum corrections are regularised by the corresponding diagram in the triangle topology, as in the second figure.}
\label{fig:exception}
\end{figure}

Finally, we consider the remaining quantum corrections that we treat perturbatively, which involve $h_{\mu\nu}$, the ghosts and matter fields. These involve attaching propagators between any pair of the $\Gamma_1$ copies in fig. \ref{fig:thirdorder}. If these propagators attach a pair that are already attached by a melonic expansion, then the resulting contribution will be UV finite thanks to the regulating factor \eqref{regulategen}. At first sight that leaves some unprotected quantum corrections, namely where we attach propagators from the left-most $\Gamma_1$ to the right-most $\Gamma_1$ in the linear topology. However it is straightforward to see that the combinatorics are such that these are regularised by the corresponding diagram in the triangle topology, as illustrated in fig. \ref{fig:exception}

\subsection{General case}

\begin{figure}[ht]
\centering
\includegraphics[scale=0.35]{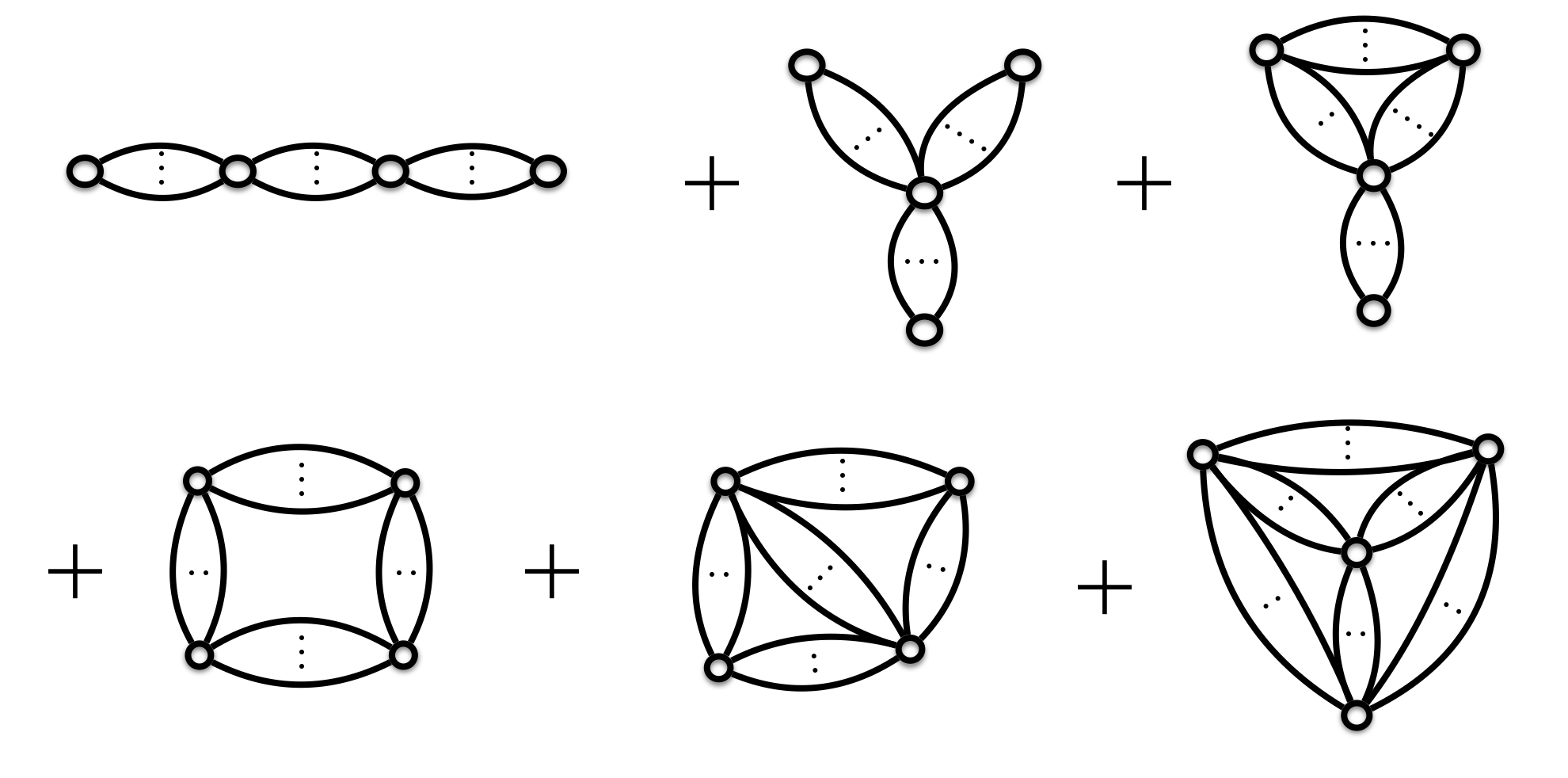}
\caption{The fourth-order part, $\Gamma_4$, joins four copies of $\Gamma_1$ together using melonic expansions, in such a way as to avoid overcounting and exclude 1PR and disconnected diagrams.}
\label{fig:fourthorder}
\end{figure}

It is clear that  we now have the broad general pattern. Let us sketch how it would work in $\Gamma_4$. Firstly, some thought makes clear that its diagrammatic expansion must take the form given fig. \ref{fig:fourthorder}. Then it is evident that the purely $\ph$ melonic expansions will provide UV regulating factors \eqref{regulategen} sufficient to ensure that no UV divergences can be produced by bringing any set of $\Gamma_1$ integrands close to each other. As before, the correction terms that correspond to disconnected or 1PR pieces in a more complicated topology will effectively reduce the corresponding diagram to simpler topologies earlier in our list, whose UV finiteness we will have already established. As a simple example, which is a little different to the previous cases, consider the linear topology in fig. \ref{fig:fourthorder} where for the middle factor we keep only the $-1$ correction. This will break the diagram into two copies of the $\Gamma_2$ melonic expansion which we have already confirmed are UV finite.

Now consider the contributions from explicit bilinear terms in the copies of $\Gamma_1$. If we retain bilinear terms such that they are connected to only a single copy of $\Gamma_1$, the result is a  tadpole again. For example this is true for the bilinear terms  in an outer $\Gamma_1$ of the linear topology, or if we retain only the bilinear terms in three of the $\Gamma_1$ copies in the square topology (the first diagram in the second line). The bilinear terms make no contribution if $\Gamma_1$ is connected to more than two other copies, for example the middle vertex in the last diagram, since these need more than two legs. 

Thus the only non-trivial contributions come from retaining the bilinear terms in a $\Gamma_1$ copy that is connected to two other $\Gamma_1$ copies. This is true for example for either of the top $\Gamma_1$ copies in the third diagram of the first line of fig. \ref{fig:fourthorder}. However these then reduce to UV finite contributions in a similar way to that in fig. \ref{fig:bilinear}. 
As a second example  consider retaining only the bilinear term in just one of the $\Gamma_1$ copies in the square topology. At first sight this looks problematic, in a similar way to the first diagram in fig. \ref{fig:exception}, because now the correction is attached to two $\Gamma_1$ copies in such a way \TRM{that} it is not directly regularised by a melonic expansion. However the combinatorics must be such that this contribution combines with the contributions from keeping only the bilinear terms in one of the two possible $\Gamma_1$ copies in the next topology (the middle diagram in the second line of fig. \ref{fig:fourthorder}). Thus the latter's diagonal melonic expansion provides the needed UV regularisation from its purely $\ph$ quantum corrections in this case.

Similar arguments to before, establish that the remaining quantum corrections, those involving the other fields and propagators passing between any pair of $\Gamma_1$, are all UV regularised. This regularisation is either immediate because those copies of $\Gamma_1$ are attached by a melonic expansion, or follows from the correction to a corresponding topology later in the list in fig. \ref{fig:fourthorder}.

\section{Compact formulae for the reorganised perturbation series}
\label{sec:reorganise}

In this section we prove the expressions for the $\Gamma_n$ that we used in sec. \ref{sec:partialresum} to demonstrate that perturbative quantum gravity can be partially resummed into a series of UV finite terms.
As in sec. \ref{sec:partialresum}, we use compact DeWitt notation. 
For the fermionic fields it is helpful to introduce both left ($\partial_l$) and right ($\partial_r$) derivatives. 
We write the partition function as 
\be \label{partfn}
\mathcal{Z}[J] = \mathrm{e}^{W[J]} = \int\!\! \mathcal{D}\phi^A\, \exp\left(-\frac12\phi^A\Delta^{-1}_{AB}\phi^B-\epsilon S_I[\phi]+J_A\phi^A\right)\,.
\ee
Now consider perturbing the propagators $\Delta^{AB}\mapsto\Delta^{AB}+\delta\Delta^{AB}$. This induces the following change to the generator of connected diagrams:
\be 
\delta W = -\frac12 \frac{\partial_lW}{\partial J_A}\delta\Delta^{-1}_{AB}\frac{\partial_lW}{\partial J_B} -\frac12 \delta\Delta^{-1}_{AB}\frac{\partial^2_lW}{\partial J_A\partial J_B}\,.
\ee
From the Legendre transform relation, $W[J] = -\Gamma[\Phi]+J_A\Phi^A$, where the $\Phi^A$ are the so-called classical fields, we then get by standard manipulations,
\be 
\label{flow}
\delta\Gamma_I[\Phi] = -\frac12\, \mathrm{Str}\left(\delta\Delta\, \Delta^{-1}\left[1+\Delta\Gamma^{(2)}_I\right]^{-1}\right)\,,
\ee
where we have introduced a functional supertrace Str$\,\mathcal{M} = (-)^A\, \mathcal{M}^{A}_{\ \,A}$, and the Hessian is given by
\be 
\label{Hessian}
\Gamma^{(2)}_{I\ AB} = \frac{\partial_l}{\partial\Phi^A}\frac{\partial_r}{\partial\Phi^B}\Gamma_I\,.
\ee
Equation \eqref{flow} takes a closely similar form to the Legendre flow equation used in Exact Renormalization Group investigations \cite{Nicoll1977, Wetterich:1992, Morris:1993}, see also  \cite{Weinberg:1976xy,Morris:2015oca,Bonini:1992vh,Ellwanger1994a,Morgan1991}. In fact we are following the notation in \cite{Igarashi:2019gkm,first,Morris:2020blt,Kellett:2020mle} from which we also will lift the closed form expressions for $\Gamma_1$ and $\Gamma_2$. 
But we emphasise that here no renormalization group is being used, nor are we introducing any novel interactions. We use equation \eqref{flow} merely as an effective way to handle the combinatorics involved in the reorganisation of the standard perturbation series for 1PI (one-particle irreducible) diagrams.

\subsection{First order}
\label{sec:reofirst}

To lowest order in $\epsilon$ we thus get,
\be \delta \Gamma_1 = \frac12\, \mathrm{Str}\left(\delta\Delta\, \Gamma^{(2)}_1\right) = \frac12\, \delta\Delta^{AB}\partial_B\partial_A\, \Gamma_1\,, \ee
where now we introduce the short-hand $\partial_A = \partial_l/\partial\Phi^A$. This is in effect a first order differential equation which is solved by an exponential of $\half\Delta^{AB}\partial_B\partial_A$ acting on a functional that has no propagators. The latter clearly has to be the classical interaction and thus:
\be 
\label{tadpoles}
\Gamma_1 = \mathrm{e}^{\half\Delta^{AB}\partial_B\partial_A}S_I[\Phi] = \mathrm{e}^{\Po/2} S_I[\Phi] \,.
\ee
This is the sum over tadpole diagrams that we used already in sec. \ref{sec:partialresum} and illustrated in fig. \ref{fig:tadpoles}. In the second equality we introduce the compact notation, 
\be \label{Po}\Po = \Delta^{AB}\partial_B\partial_A\,.\ee 
In this form, equation \eqref{tadpoles} is the one given in slightly different notation in eqn. \eqref{Gone}.

\subsection{Second order}
\label{sec:reosecond}

To second order in $\epsilon$ we have to solve the inhomogeneous differential equation:
\be 
\label{ode2}
\delta\Gamma_2 -\frac12\,\mathrm{Str}\left(\delta\Delta\, \Gamma^{(2)}_2\right) =-\frac12\,\mathrm{Str}\left(\delta\Delta\,\Gamma^{(2)}_1\Delta\Gamma^{(2)}_1\right)\,.\ee
This can be solved by introducing the appropriate integrating factor, equivalently by defining
\be 
\label{G2stripped}
\Gamma_2 = \mathrm{e}^{\Po/2}\,\sGamma_2[\Phi]\,,
\ee
where $\sGamma_2$ is the second-order functional stripped of most of its propagators. Substituting \eqref{G2stripped} into \eqref{ode2} and rearranging gives
\be 
\label{sGa2eq}
\delta\sGamma_2 = -\frac12\,\mathrm{e}^{-\Po/2}\, \mathrm{Str}\left(\delta\Delta\,\Gamma^{(2)}_1\Delta\Gamma^{(2)}_1\right)
= -\frac12\,\delta\Po_{12}\,\Po_{12}\,\mathrm{e}^{-\Po_{12}}S_I\,S_I\,.
\ee
In the second equality we use the Leibniz rule to write  $\partial_A = \partial^1_A+\partial^2_A$, where $\partial^1_A$ acts only on the first $\Gamma_1$ and $\partial^2_A$ acts only on the second. This means that 
\be \label{Poexpand}\Po = \Po_{11}+\Po_{22}+2\Po_{12}\,,\ee 
where we are now using the notation in eqn \eqref{Poij}.
Substituting the tadpole solution \eqref{tadpoles} for the two instances of $\Gamma_1$, and expanding out the supertrace, then completes \eqref{sGa2eq}. Equation \eqref{sGa2eq} can now be solved by integration by parts, since it is analogous to finding the function $f(x)$ that satisfies $f'(x)=-\half x\,\mathrm{e}^{-x}$. Thus we find
\be \sGamma_2[\Phi] = \frac12\,(\Po_{12}+1)\,\mathrm{e}^{-\Po_{12}}S_I^{\,2} + C_2[\Phi]\,,\ee
where $C_2$ is second order in $S_I$ but has no propagators. However if we set the propagators to zero in the first term and also in \eqref{G2stripped}, the result for $\Gamma_2$ must be 1PI. That tells us that in fact $C_2=-\half S_I^{\,2}$. Now, premultiplying by $\mathrm{e}^{\Po/2}$ to convert the above into $\Gamma_2$, and expanding $\Po$ as in \eqref{Poexpand} again, gives our final result:
\be \Gamma_2[\Phi] = -\frac12 \left(\mathrm{e}^{\Po_{12}}-1-\Po_{12}\right) \Gamma_1^{\,2}\,,\ee
\ie the melonic expansion formula \eqref{melons} that we already used in sec. \ref{sec:partialresum} and which was illustrated in fig. \ref{fig:melons}. 

\subsection{Third order}
\label{sec:reothird}

This method generalises to any order, although the algebra of course gets progressively more complicated. To illustrate, we sketch the steps for $\Gamma_3$. From \eqref{flow} we have that
\be \delta\Gamma_3 = \frac12\, \mathrm{Str}\left( \delta\Delta\, \Gamma^{(2)}_3 - \delta\Delta\,\Gamma^{(2)}_1\Delta\Gamma^{(2)}_2 -\delta\Delta\,\Gamma^{(2)}_2\Delta\Gamma^{(2)}_1 + \delta\Delta\,\Gamma^{(2)}_1\Delta\Gamma^{(2)}_1\Delta\Gamma^{(2)}_1\right)\,.\ee
Using \eqref{Poij} this can be rewritten as
\be \delta\Gamma_3 -\frac12\delta\Po\, \Gamma_3 = -\delta\Po_{12}\,\Po_{12}\, \Gamma_1\, \Gamma_2 +\frac12 \delta\Po_{31}\,\Po_{12}\Po_{23}\,\Gamma_1^{\,3}\,.\ee
Substituting \eqref{melons} and symmetrising gives:
\be \delta\Gamma_3 -\frac12\delta\Po\, \Gamma_3 = \frac12\left\{ \left(\delta\Po_{12}+\delta\Po_{13}\right)\left(\Po_{12}+\Po_{13}\right)\left(\mathrm{e}^{\Po_{23}}-1-\Po_{23}\right)+\delta\Po_{31}\,\Po_{12}\Po_{23}\right\} \Gamma_1^{\,3}\,.\ee
Now premultiplying by $\mathrm{e}^{-\Po/2}$ to form $\sGamma_3$, analogous to \eqref{G2stripped}, and on the right hand side using \eqref{tadpoles} again  and Liebnitz, $\Po = \sum_{i,j=1}^3\Po_{ij}$, puts the equation in a form where again it can be integrated by parts, starting this time with the factor in front of the exponential that is a cubic in $\Po_{ij}$. After symmetrisation this cubic term can be put in the form 
\be -\frac12\left(\delta\Po_{12}+\delta\Po_{23}+\delta\Po_{31}\right)\Po_{31}\Po_{23}\,\mathrm{e}^{-\Po_{12}-\Po_{23}-\Po_{31}}\Gamma_1^{\,3}\,,\ee
which can be combined with the exponential 
so that integrating by parts leaves only a quadratic factor. Continuing in this way leads to the final answer, where again the integration `constant' was determined by ensuring that the result is 1PI:
\be \Gamma_3 = \left\{ \frac16 \,\mathrm{e}^{\Po_{12}+\Po_{23}+\Po_{31}}-\Po_{12}\,\mathrm{e}^{\Po_{23}}-\frac12\,\mathrm{e}^{\Po_{12}}+\frac12\Po_{13}\Po_{23}+\Po_{12}+\frac13\right\}\Gamma_1^{\,3}\,.\ee
After some manipulation this can be put in the more intuitive form of eqn. \eqref{thirdorder},
as illustrated diagrammatically in fig. \ref{fig:thirdorder}.

\section{BRST invariance and the Zinn-Justin identity}
\label{sec:BRST}

The BRST invariance for the quantum fields was already defined in equations (\ref{Qg},\ref{Qghost}), with those for $\ph$ and $h_{\mu\nu}$ defined implicitly through the action of $Q$ on the total metric $g_{\mu\nu}$. Substituting the latter's parametrisation \eqref{paramg}, into \eqref{Qg}, 
and contracting with $\hat{g}^{\mu\nu}$ gives the BRST transformation of $\ph$ in explicit form:
\be \label{Qph} Q\ph = \partial_\mu c^\mu +\kappa c^\alpha\partial_\alpha \ph\,,  \ee
and thus from the parametrisation \eqref{paramg} we also obtain that for $\hat{g}_{\mu\nu}$:
\be \label{Qgh} Q\hat{g}_{\mu\nu}  = 2\kappa\partial_{(\mu} c^\alpha \hat{g}_{\nu)\alpha} -2\frac{\kappa}{d}\hat{g}_{\mu\nu}\partial_\alpha c^\alpha +\kappa c^\alpha\partial_\alpha\hat{g}_{\mu\nu}\,. \ee
This defines implicitly the BRST transformation of $h_{\mu\nu}$. It of course has to remain implicitly defined until we fully specify the parametrisation of $\hat{g}_{\mu\nu}$ (parametrised in \eqref{paramhg} to first order), but we note that it is already clear from this equation that $Qh_{\mu\nu}$ is not protected by a $\ph$ exponential.

\TRM{That is fine since it is only interactions, not the BRST transformations themselves, that need to be protected by $\ph$ exponentials, but} to test  for BRST invariance on the Legendre effective action, $\Gamma$, we need the action of BRST on the classical fields $\Phi^A$ and these are determined by the Zinn-Justin identity. To derive this identity, we need to add to the action the source terms
\be - \int\!\! d^dx \,\Big\{ \ph^*\, Q\ph+ h^{*\mu\nu}\, Qh_{\mu\nu} + c^*_\mu\, Qc^\mu +\bar{c}^*_\mu\, Q\bar{c}^\mu \Big\}\,, \ee
where $\Phi^*_A =$ $\ph^*$, $h^{*\mu\nu}$, $c^*_\mu$, and $\bar{c}^*_\mu$, are sources for the BRST transformations acting on the corresponding quantum fields. At this point we encounter a problem because the $\ph^*$ and $h^{*\mu\nu}$ terms introduce interactions that are not protected by positive exponentials of $\ph$, and thus these terms get quantum corrections that are not UV finite.\footnote{Notice from \eqref{Qghost} that $Q\bar{c}^\mu$ is protected by a $\ph$ exponential. Although $Qc^\mu$ is not, it can be shown that the short distance divergences it can cause are actually harmless (\ie integrable).} 

This problem can be solved by choosing a safer \TRM{if slightly more involved} form for the BRST transformation, taking it instead to be the Lie derivative along $\kappa\,\mathrm{e}^{\kappa\ph} c^\mu$. Since this amounts to replacing $c^\mu$ with $\mathrm{e}^{\kappa\ph} c^\mu$, it is clear from equations (\ref{Qph},\ref{Qgh}) that the $\ph^*$ and $h^{*\mu\nu}$ terms are then protected. Making this substitution on \eqref{Qghost} and using the substituted form of \eqref{Qph} we get a fully protected $c^*$ term, since:
\be Qc^\mu = \kappa\, \mathrm{e}^{\kappa\ph}\left( c^\nu\partial_\nu c^\mu - \partial_\nu c^\nu c^\mu -\kappa c^\nu \partial_\nu\ph\, c^\mu\right)\,.\ee
The $\bar{c}^*$ term is even better protected. In fact in this case we could use the standard gauge fixing functional $F_\mu = \delta^{\alpha\beta}\partial_\alpha h_{\beta\mu}-\left(\frac2d-1\right)\partial_\mu\ph$, \ie without the protective interactions in \eqref{F}, since from \eqref{flatgf} the gauge fixing term then no longer has interactions, while the ghost action and the $c^*$ term now receive their protection from the positive exponential of $\ph$ appearing in the $Q\ph$ and $Qh_{\mu\nu}$ transformations.

Now it is straightforward to derive the Zinn-Justin identity following the standard steps (here for the on-shell BRST version). Applying the BRST charge to the action plus source terms we get, by invariance of the measure, 
\be \left( J_A \frac{\partial}{\partial\Phi^*_A} +\alpha \kappa \delta^{\mu\nu}\bar{\eta}_\mu \bar{c}^*_\nu \right) \mathcal{Z}[J,\Phi^*] = 0\,,\ee
where $\bar{\eta}_\mu$ is the source term for $\bar{c}^\mu$ (its explicit appearance arising from the fact that $Q^2\bar{c}^\mu$ is proportional to the $\bar{c}$ equation of motion and thus can be absorbed by a shift of $\bar{c}$ integration variable), and thus
\be \frac{\partial\Gamma}{\partial\Phi^A}\frac{\partial\Gamma}{\partial\Phi^*_A}-\alpha \kappa \delta^{\mu\nu}\bar{c}^*_\mu \frac{\partial\Gamma}{\partial\bar{C}^\mu} = 0\ee
(where $\bar{C}$ is the classical antighost field).

The Zinn-Justin identity can now be expanded as an infinite series in $\epsilon$, \ie over the $\Gamma_n$ terms, each of which are UV finite and which now include (UV finite) interactions with the $\Phi^*$ source fields. Unfortunately, as we will see in the next sections, 
we then encounter a more profound difficulty in that no finite set of the resulting $\Gamma_n$ will satisfy the Zinn-Justin identity by themselves. 

\section{UV finiteness on a curved background}
\label{sec:curved}

So far we have expanded around flat background. Now let us show that these properties can be extended to curved backgrounds and the background field method.  It is straightforward to extend sec. \ref{sec:two} so that the two key properties are preserved. However as we will see, in order to realise both background diffeomorphism invariance and UV finiteness, we will need to sum over most if not all of the corrections contained in the $\Gamma_n$ of the $\epsilon$ expansion \eqref{epsexp}.

Expanding the metric about a background metric $\bar{g}_{\mu\nu}$ we now write
\be g_{\mu\nu} = \bar{g}_{\mu\nu}+\kappa H_{\mu\nu} +O(\kappa^2)\,,\ee
splitting out the trace of the fluctuation field, by writing
\be\label{traceb} H_{\mu\nu} = h_{\mu\nu}+\tfrac2d \,\ph\,\bar{g}_{\mu\nu}\,,\qquad\text{where}\quad \ph =\half \bar{g}^{\mu\nu}H_{\mu\nu}\,,\quad\text{and}\quad \bar{g}^{\mu\nu}h_{\mu\nu} =0\,.\ee
Furthermore for perturbative purposes we define the background fluctuation field via
\be\label{bkg} \bar{g}_{\mu\nu} =\delta_{\mu\nu}\TRM{+}\kappa \bar{H}_{\mu\nu} +O(\kappa^2)\,.\ee
Non-perturbatively in $\kappa$, we still parametrise the metric with an exponential of $\ph$ and $\hat{g}_{\mu\nu}$ for the rest, as in \eqref{paramg},
but now $\hat{g}_{\mu\nu}$  is a function of both $h_{\mu\nu}$ and $\bar{g}_{\mu\nu}$, and to be consistent with above must satisfy
\be \hat{g}_{\mu\nu} = \bar{g}_{\mu\nu}+\kappa h_{\mu\nu} + O(\kappa^2)\,.\ee
Since the parametrisation of the metric takes the same form, the Einstein-Hilbert action still takes the same form, \ie as in \eqref{EHreparam}. However we need to replace the gauge fixing and ghost terms by ones that are manifestly background diffeomorphism invariant so, in minimal-coupling fashion, we write
\be\label{backgf}\int\!\! d^dx \,\sqrt{\bar{g}}\left\{ \frac{\alpha}{2}\bar{g}^{\mu\nu}F_\mu F_\nu -\frac{1}{\kappa}\bar{c}^\mu QF_\mu\right\}\,, \ee
and
\beal F_\mu &= \frac1\kappa \bar{g}^{\alpha\beta} \left(\bar{\nabla}_\alpha g_{\beta\mu}-\frac12\bar{\nabla}_\mu g_{\alpha\beta} \right) =\bar{g}^{\alpha\beta} \bar{\nabla}_\alpha h_{\beta\mu}-\left(\frac2d-1\right)\bar{\nabla}_\mu\ph +O(\kappa)\nn\\
&= \bar{g}^{\alpha\beta} \mathrm{e}^{2\kappa \ph/d} \left(\frac1\kappa\bar{\nabla}_\alpha \hat{g}_{\beta\mu}-\frac1{2\kappa}\bar{\nabla}_\mu \hat{g}_{\alpha\beta} +\frac2d\TRM{\bar{\nabla}}_\alpha\ph\,\hat{g}_{\beta\mu}-\frac1d\bar{\nabla}_\mu\ph\,\hat{g}_{\alpha\beta}\right)\,.\label{backF}
\eeal
The BRST transformations for $g_{\mu\nu}$ and $c^\mu$ remain as before because they are formed from Lie derivatives. As is well known, they are equal to the versions where the partial derivatives are replaced by (background) covariant derivatives (thus making the formulae manifestly background diffeomorphism invariant). Only the $\bar{c}$ BRST transformation changes, namely to $Q\bar{c}^\mu = \alpha \bar{g}^{\mu\nu}F_\nu$.

Working on a curved background has introduced new interactions, which however are also all protected by positive exponentials of $\ph$. Therefore, we have automatically also the finiteness property we established in secs. \ref{sec:partialresum} and \ref{sec:reorganise}.
It is also clear that we can further protect the BRST transformations as we did in sec. \ref{sec:BRST} in order to formulate the Zinn-Justin identity in the presence of a background metric.  

As we remarked in sec. \ref{sec:two} when considering the flat space version, we can introduce an auxiliary field $b^\mu$ to get full off-shell BRST invariance if preferred. However, following ref. \cite{Mandric:2023dmx} this would at first sight appear to present a problem now, because the bilinear $b$ term we need, namely $\sqrt{\bar{g}}\bar{g}_{\mu\nu}b^\mu b^\nu$, has interactions with the background metric which are not protected by a positive exponential of $\ph$. However this is not the case. The $b$ field only appears in internal lines of 1PI diagrams and thus, as shown in ref. \cite{Mandric:2023dmx},  the net result of such quantum corrections is the same as one gets from integrating out the $b$ field. But doing that, takes us back to the on-shell formulation considered here which we have just shown is fully protected.

Although we can thus again perform a partial resummation of the perturbation theory into the analogous series of UV finite terms, $\Gamma_n$, we encounter a problem in that these do not individually respect background diffeomorphism invariance. We can establish this directly by demonstrating that the $\Gamma_n$ do not satisfy the appropriate Ward identities. At the end of this section and in sec. \ref{sec:limits} we will consider the consequences from this point of view. However another way to illuminate the problem is instead to explain why the obvious manifestly background diffeomorphism invariant generalisation will fail, namely the generalisation 
where we work non-perturbatively in the background by using covariant propagators formed from the inverse of the covariant d'Alembertian $\bar{\Box}=\bar{\nabla}^\mu\bar{\nabla}_\mu$ and its appropriate generalisations for vector and tensor fields. 

\begin{figure}[ht]
\centering
\includegraphics[scale=0.35]{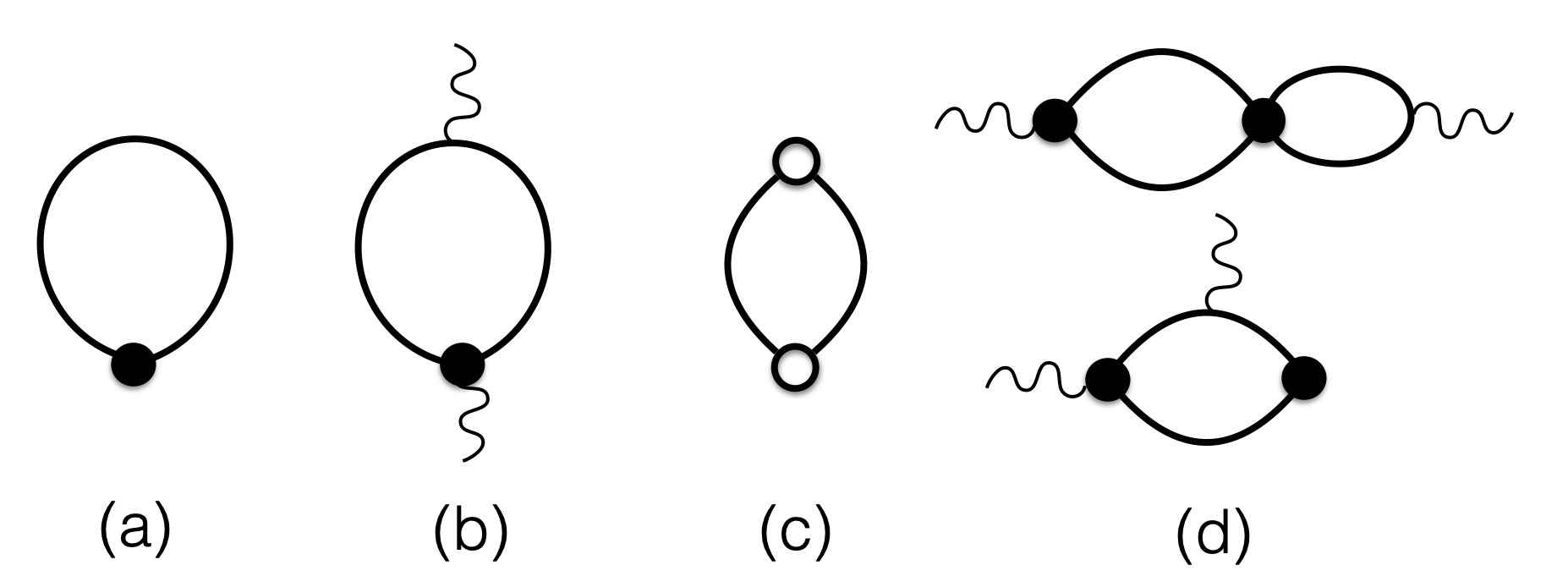}
\caption{When covariantised, the one-loop tadpole diagram (a) from the tadpole expansion of fig. \ref{fig:tadpoles}, contains the self-energy diagram (b). In (b), the external legs are explicitly drawn and represent the background fluctuation field $\bar{H}_{\mu\nu}$, the bottom interaction coming from the classical action vertex and the top from the covariantised propagator. This self-energy diagram is in fact contained in diagram (c) which is the first correction from the melonic expansion of $\Gamma_2$ in fig. \ref{fig:melons}. However this in turn is covariantised and thus contains diagrams such as those labelled (d).}
\label{fig:covariant}
\end{figure}

If we covariantise in this way, the $\Po_{ij}$ in eqn \eqref{Poij} will then be invariant under (background) diffeomorphisms and thus it is straightforward to generalise the partial resummations we had previously in such a way that they are now background covariant. 
However these resummations are not the ones we want, because this generalisation  fails to be UV finite. Firstly, the terms bilinear in the quantum fields that we need to subtract to form the vertices, such as those in \eqref{EHvertex} for the Einstein-Hilbert action, are now covariantised, and thus contain interactions with the background metric. Therefore these terms can no longer be discarded in the way we did at certain intermediate steps in sec. \ref{sec:partialresum} and furthermore since these covariantisations of bilinear terms are not protected by a $\ph$ exponential, they lead to unregularised UV divergences. Unregularised UV divergences also appear in the resummations, for example this happens already at the first order, \ie in the summation over tadpoles, \viz fig.  \ref{fig:tadpoles} and eqn. \eqref{Gone}. Since the propagators are continuously interacting with the background, the corrections in fig. \ref{fig:tadpoles} are no longer tadpole integrals that can be set to zero, but contain important corrections which explore the interaction of the quantum fields with the background metric. For example the one-loop correction in fig. \ref{fig:tadpoles}, now contains the self-energy diagram, as illustrated in the first two diagrams of fig. \ref{fig:covariant}. 

Clearly the reason the partial resummations are now failing is because we have included the interactions with the background, which of course were those in the original classical action, but without also incorporating the melonic expansions over the $\ph$ corrections which lead to UV regularisation. Thus we are forced instead to keep the subtracted bilinear terms truly bilinear, leaving all interactions with the background inside vertices that are protected by a $\ph$ exponential (such as the first term in \eqref{EHvertex}). Continuing with our tadpoles example, we note that the background self-energy diagram is then in fact contained in diagram (c) of fig. \ref{fig:covariant}. This is the first diagram in the second order correction $\Gamma_2$, \ie the melonic expansion of fig. \ref{fig:melons}. To incorporate the UV regularisation we then need to add the remaining diagrams of fig. \ref{fig:melons}, \ie we need to add the full $\Gamma_2$ correction.

Now, for $\Gamma_2$ to be fully covariant to second order in $\bar{H}_{\mu\nu}$, we need to add to its copies of $\Gamma_1$ the terms that covariantise their  tadpole corrections, \eg the diagrammatic contributions (a) of fig. \ref{fig:covariant}. For example we need to incorporate the upper correction of (d) in fig. \ref{fig:covariant}. This correction is supplied by the linear topology of $\Gamma_3$, as in fig. \ref{fig:thirdorder}, and it is UV regularised by including the full melonic expansions for that topology.

We also need covariantisations corresponding to the lower diagram of fig. \ref{fig:covariant} (d), where a background interaction covariantises one of the propagators in the one-loop melonic contribution. This diagram is supplied by the triangle topology of $\Gamma_3$, as in fig. \ref{fig:thirdorder}, and once again, in order to incorporate the UV regularisation we need to include the full set of melonic corrections for this topology. 

Proceeding in this way, we see that we end up having to sum over many, if not all, of the UV finite correction terms we previously derived. For example from $\Gamma_4$, we  need the linear topology of fig. \ref{fig:fourthorder}, as is clear from a similar argument that led to requiring the linear topology of $\Gamma_3$. The next $\Gamma_4$ topology in fig. \ref{fig:fourthorder} is also needed, because it is involved in regularising the covariantised tadpole corrections of the middle vertex in the linear topology of $\Gamma_3$, as in fig. \ref{fig:thirdorder}. The third $\Gamma_4$ topology in fig. \ref{fig:fourthorder} can be seen to be needed to regularise the covariantised propagators in the linear topology of $\Gamma_3$, and the fifth $\Gamma_4$ topology is needed to regularise the covariantised propagators in the triangle topology of $\Gamma_3$.

To summarise, an appropriate gauge fixing can be found such that perturbative quantum gravity about a curved background metric, can also be partially resummed into a series of UV finite correction terms, $\Gamma_n$. However the $\Gamma_n$ are not individually background diffeomorphism invariant. In order to recover background diffeomorphism invariance, it is clear that we would have to sum over many, if not all, of the (parts of) $\Gamma_n$. 

In other words the Ward identities expressing background diffeomorphism invariance are  broken at any finite order in the $\epsilon$ expansion.
Although we have not explicitly addressed BRST invariance, \ie the Zinn-Justin identity, it  leads to analogous but more involved Ward identities, and these clearly will also be violated at any finite order. Technically, in both cases, the Ward identities relate a  space-time differential $\partial_\mu$ contracted into higher-point vertices, to lower-point vertices. Working strictly order by order in $\kappa$, these relations are exactly satisfied. Here however they are broken term by term by our partial resummation. Specifically, the breaking arises because $\partial_\mu$ also acts on the UV regulating exponential \eqref{regulategen}. Thus violations of the Ward identities result from factors of the form
\be \cdots\, \xi^2\beta_{\alpha_i}\beta_{\alpha_j}\kappa^2\partial_\mu\Delta(x_i-x_j) \exp\big\{-\xi^2\beta_{\alpha_i}\beta_{\alpha_j}\kappa^2\Delta(x_i-x_j)\big\}\,\cdots\, \ee
(where we have included the errant exponential and the rest of the term is represented by the ellipses). Now the problem is that regions of integration where the UV regularisation term is needed, correspond to places where $\Delta(x_i-x_j)$ is large. The exponential effectively cuts off these integrals over $x_i$ when 
\be \xi^2\beta_{\alpha_i}\beta_{\alpha_j}\kappa^2\Delta(x_i-x_j)\sim O(1)\,,\ee
and thus we see that violations of the Ward identity are always of the same size as the terms themselves.

\section{Limiting cases} 
\label{sec:limits}

Clearly this problem is also related to the fact that we no longer have a small parameter to control the expansion. One might hope that one could use the spacetime dimension $d$ in the passage to some limit (\eg $d\to\infty$ \cite{Strominger:1981jg,Bjerrum-Bohr:2003veq,Hamber:2005vc}) to recover a suitable small parameter, but unfortunately this does not appear to help. 

At  first sight we do have a parameter to control the $\Gamma_n$, namely the gauge parameter $\xi$ that weights the UV regulating exponential $\exp\big\{-\xi^2\beta_{\alpha_i}\beta_{\alpha_j}\kappa^2\Delta(x_i-x_j)\big\}$, \cf \eqref{regulategen}. ($\xi$ also appears in some multiplying factors, see \eg the example contribution \eqref{Gtwoexample}.) Indeed by sending $\alpha\to (d-2)/(d-1)$ from above, we can make $\xi$ arbitrarily small whilst leaving finite the other propagators and the vertices (see (\ref{hpalph}--\ref{hhalph}) and footnote \ref{foot:xi}). By choosing  $F_\mu$ such that at the linearised level it is the most general gauge:
\be F_\mu = \frac1\kappa \TRM{\bar{g}}^{\alpha\beta} \left(\alpha\TRM{\bar{\nabla}}_\alpha g_{\beta\mu}+\beta\TRM{\bar{\nabla}}_\mu g_{\alpha\beta} \right) \,,\ee
where now the gauge fixing and ghost terms are:\footnote{We return to the previous gauge (\ref{flatgf},\ref{F}) by setting $\beta=-\frac12\alpha$ and then redefining $\alpha^2 \mapsto \alpha/2$.}
\be\label{gengf}\int\!\! d^dx \,\left\{ \TRM{\bar{g}}^{\mu\nu}F_\mu F_\nu -\frac{1}{\kappa}\bar{c}^\mu QF_\mu\right\}\,, \ee
we can make $\xi$ arbitrarily small and still be left with a general set of gauges in order to test gauge parameter independence. Indeed, since in this case we find \cite{Capper:1979ej}:
\beal
\label{ppgen}
\langle \ph(p)\,\ph(-p)\rangle &= -\frac{2(d-1)\alpha^2-(d-2)}{8(d-2)(\alpha+\beta)^2p^2}  = -\frac{\xi^2}{p^2}\,,
\\
\label{hpgen}
\langle h_{\mu\nu}(p)\, \ph(-p)\rangle &= \langle \ph(p)\,h_{\mu\nu}(-p)\rangle =
-\frac{d-2+2\alpha(\alpha+\beta d)}{4(d-2)(\alpha+\beta)^2} \left( \frac{\delta_{\mu\nu}}{d}-\frac{p_\mu p_\nu}{p^2}\right) \frac1{p^2}\,,\\
\langle h_{\mu\nu}(p)\,h_{\alpha\beta}(-p)\rangle &= \frac{\delta_{\mu(\alpha}\delta_{\beta)\nu}}{p^2} 
+2\left(\frac1{\alpha^2}-1\right)\frac{p_{(\mu}\delta_{\nu)(\alpha}p_{\beta)}}{p^4}+\frac{(d-2)(1-2\alpha^2)-2(\beta d+[d-1]\alpha)^2}{2d^2(d-2)(\alpha+\beta)^2} \frac{\delta_{\mu\nu}\delta_{\alpha\beta}}{p^2}\nn\\
&\quad +\frac{4d\beta(\alpha+\beta)+2(d-1)\alpha^2-(d-2)}{2d(d-2)(\alpha+\beta)^2}\frac{\delta_{\alpha\beta}p_\mu p_\nu+p_\alpha p_\beta \delta_{\mu\nu}}{p^4}\nn\\
&\quad\quad+\frac{(\alpha+2\beta)(2[d-3]\alpha^3-4\beta\alpha^2-[d-2][3\alpha+2\beta])}{2(d-2)\alpha^2(\alpha+\beta)^2}\frac{p_\mu p_\nu p_\alpha p_\beta}{p^6}\,,\label{hhgen}\\
\langle c^\mu(p)\,\bar{c}^\nu(-p)\rangle &= \left(\delta^{\mu\nu}-\frac{\alpha+2\beta}{2(\alpha+\beta)} \frac{p^\mu p^\nu}{p^2}\right)\frac1{p^2}\,,\label{ghgen}
\eeal
we have 
\be \xi = \frac1{2|\alpha+\beta|}\sqrt{\frac{d-1}{d-2}\alpha^2-\frac12} \ ,\ee
and $\xi$ is made arbitrarily small by sending 
\be \alpha\to \sqrt{\frac{d-2}{2(d-1)}} \ee
from above, whilst leaving $\beta$ as a free gauge parameter and leaving the other propagators finite (provided only that $\beta\ne-\alpha$). 

We can also take the limit in which $\xi$ becomes arbitrary large, provided we also allow $\langle h_{\mu\nu}\ph\rangle$ and $\langle c^\mu \bar{c}^\nu\rangle$ to become large, whilst nevertheless leaving the $\langle h_{\mu\nu} h_{\alpha\beta} \rangle$ propagator finite. Setting
\beal
\label{ablim}
\alpha &=\ \sqrt{\frac{d-2}{2(d-1)}}\left(1+\frac{d^2\varepsilon^2}{2(d-1)(d-2)}+\frac{(d^2-\gamma)d^2\varepsilon^4}{8(d-1)^2(d-2)^2}\right)\,,\nn\\
\beta &= -\sqrt{\frac{d-2}{2(d-1)}}\left(1+\frac{d\,\varepsilon^2}{2(d-1)(d-2)}+\frac{d(2d^2-[d+1]\gamma)\varepsilon^4}{16(d-1)^2(d-2)^2}\right)\,,
\eeal
where $\gamma$ is a finite free gauge parameter, and $\varepsilon\ll1$, yields $\xi^2 = 1/\varepsilon^2+O(\varepsilon^2)$, together with:
\beal
\label{pplim}
\langle \ph(p)\,\ph(-p)\rangle &= -\left(\frac1{\varepsilon^2}+O(\varepsilon^2)\right)\frac{1}{p^2}=-\frac{\xi^2}{p^2}\,,
\\
\label{hplim}
\langle h_{\mu\nu}(p)\, \ph(-p)\rangle &= \langle \ph(p)\,h_{\mu\nu}(-p)\rangle =
\left(\frac1{\varepsilon^2}-\frac{2d^2+\gamma}{8(d-1)(d-2)}+O(\varepsilon^2)\right)\left( \frac{\delta_{\mu\nu}}{d}-\frac{p_\mu p_\nu}{p^2}\right) \frac1{p^2}\,,\\
\langle h_{\mu\nu}(p)\,h_{\alpha\beta}(-p)\rangle &= \frac{\delta_{\mu(\alpha}\delta_{\beta)\nu}}{p^2} 
+\frac{2d}{d-2}\frac{p_{(\mu}\delta_{\nu)(\alpha}p_{\beta)}}{p^4}+\frac{\gamma-2d^2(d-2)}{2d^2(d-1)(d-2)} \frac{\delta_{\mu\nu}\delta_{\alpha\beta}}{p^2}\label{hhlim}\\
&\quad +\frac{2d(d-2)-\gamma}{2d(d-1)(d-2)}\frac{\delta_{\alpha\beta}p_\mu p_\nu+p_\alpha p_\beta \delta_{\mu\nu}}{p^4}+\frac{\gamma-2d(3d-4)}{2(d-1)(d-2)}\frac{p_\mu p_\nu p_\alpha p_\beta}{p^6}+O(\varepsilon^2)\,,\nn\\
\langle c^\mu(p)\,\bar{c}^\nu(-p)\rangle &= \frac{\delta^{\mu\nu}}{p^2}+\left(\frac{2(d-2)}{d\varepsilon^2}-1+\frac{\gamma}{4d}+O(\varepsilon^2)\right) \frac{p^\mu p^\nu}{p^4}\,.\label{ghlim}
\eeal

Obviously, taking the limit of small or large $\xi$ will control the size of the $\Gamma_n$ through the weighting of the regulating exponential \eqref{regulategen}, however we know beforehand that neither limit can recast the expansion over $\Gamma_n$ as a genuine perturbative expansion. We know this for example, precisely because their size is then gauge parameter dependent, but properly formulated the result for any physical quantity must be independent of the choice of gauge. Nevertheless it is instructive to see how the $\Gamma_n$ behave in these limits.

To be more explicit consider computing for pure gravity, the  $\bar{H}_{\mu\nu}$ two-point vertex. The first contribution to this will use $\Gamma_2$, but also build on it by attaching perturbatively other propagators involving $h_{\mu\nu}$ and the ghosts. There are also higher order contributions from the $\epsilon$ expansion, for example at third order one gets contributions from the triangle topology, \cf fig. \ref{fig:thirdorder}, and from the linear topology with an $\bar{H}$ vertex at each end (but not the other alternatives, such as having an $\bar{H}$ vertex in the middle, since if an outer vertex has no external field it forms a massless tadpole correction to the inner vertex, which vanishes in dimensional regularisation). 

One contribution from $\Gamma_2$ arises from using two copies of the Einstein-Hilbert vertex \eqref{EHvertex}, \ie from building on the term \eqref{Gtwoexample}. We also have contributions built on the gauge fixing and ghost terms \eqref{backgf}, but the former is already sufficient  to see the general pattern. We extract the background two-point vertex by taking the functional derivative with respect to $\bar{H}_{\mu\nu}(x)$ and $\bar{H}_{\alpha\beta}(y)$ and then setting all fields to zero. This will remove the $x$ and $y$ integrals in \eqref{EHvertex}, but on transferring to momentum space we will be left with an integral over $x-y$, including a factor of $\mathrm{e}^{ip\cdot(x-y)}$. The spacetime differentials in \eqref{Gtwoexample} either get directly converted to $p^\mu$ by acting on this exponential, or differentiate a propagator thus increasing the power of $1/|x-y|^2$, and/or leave us with vectors $(x-y)^\mu$ in the numerator which we can trade for $\partial/\partial p_\mu$. 

In this way, relabelling the integration variable, we can reduce all the $\Gamma_2$ two-point spacetime integrals to terms of the form 
\be\label{Ik} I_k = \int\!\!d^4x \, \frac1{x^{2k}}\, \mathrm{e}^{-\xi^2\beta^2\kappa^2\Delta(x)+ip\cdot x}\,, \ee
for some non-negative integer power $k$. Here we have specialised to the most interesting case of $d=4$ dimensions.\footnote{After some manipulation the $I_k$ can in this case be evaluated in closed form in terms of Meijer-$G$ functions, which would thus enable further development for example analytical continuation back to Minkowski signature.}
By \eqref{prop} the $1/x^{2k}$ power is then, up to a numerical factor, just a product of $k$ scalar propagators. Thus, without the UV regulating exponential, the $I_k$ are the simplest melonic scalar $(k-1)$-loop self-energy diagrams (formed from two $(k+1)$-point vertices, compare fig. \ref{fig:melons}). Thanks to the regulating exponential they are all UV finite, and thanks to the Fourier factor they are all IR finite provided $p\ne0$. By dimensions these $I_k$ are multiplied by a term  $\propto\kappa^{m}p^{6+m-2k}$, where $m$ is a non-negative integer and the power of $p$ is schematic for some tensor structure of this dimension.

Considering now $I_2$, we note that it is a function only of $\xi\beta\kappa p$. It has a log singularity in the small $\xi$ limit, with a coefficient which can be determined by evaluating $\xi\partial_\xi I_2$ in the limit $p\to0$. Thus we find that 
\be\label{sing} I_2=-\frac{\pi^2}2\ln(\xi\beta\kappa p)+\cdots\,, \ee 
where the ellipses stand for corrections that are either finite or vanish as $\xi\beta\kappa p\to0$. This can be compared to the integral without the UV regulating exponential, which is divergent, but has the same $\ln p$ dependence, \viz $-\frac{\pi^2}2\ln(p/\mu)$, as follows from \eg dimensional regularisation. Thus, as expected, we see that for small $\xi$ the integral goes back to that of standard perturbation theory, except that it is effectively cut off in the UV by 
\be \frac1{\xi\beta\kappa} = \frac{M}{2\xi\beta}\,,\ee
$M$ being the reduced Planck mass. This is the pattern one sees in the other integrals $I_k$. From \eqref{Ik}, their leading singularity in small $\xi$ limit can be found by differentiating or integrating \eqref{sing} with respect to $\xi^2\beta^2\kappa^2$, \ie with respect to the regulating scale.

The problem is that with $1/\xi\kappa$ playing the r\^ole of the cutoff, orders of perturbation theory in $\kappa$, and regularised-divergences, get mixed up with each other, and we have no way of disentangling them. In particular, higher orders in perturbation theory can contribute the same amount by supplying higher powers of $\kappa$ from the Feynman rules but also higher powers of $1/\kappa$ from the regularised UV divergences. Meanwhile differentiated $\ph$ propagators supply extra factors of $\xi$, as appear for example in \eqref{Gtwoexample}, so that powers of $\xi$ do not allow the corrections to be organised either. One also confirms in this way that the individual $\Gamma_n$ cannot be gauge invariant. For example, the term  in \eqref{Gtwoexample} with four differentiated propagators (that appears in the third line), contributes amongst other things a contribution of the form $\sim \kappa^6\beta^4\xi^8 I_6 $ whose leading behaviour in the small $\xi$ limit goes as $1/\kappa^2$, providing the graviton with a Planck sized `mass' term. This would be cancelled by higher orders in $\epsilon$ if there exists an appropriate resummation of the $\Gamma_n$ that recovers gauge invariance, but we do not know how to do this. 

In the other limit, in which $\xi$ is taken large, $\langle h_{\mu\nu}\ph\rangle$ is also large as we have already noted, \cf eqn. \eqref{hplim}. However it is not a problem to sum over these types of correction as well. The handful of instances of  $\partial_\mu\ph$ yield some factors of differentiated propagators, analogous to those in \eqref{Gtwoexample}, whilst  $\exp\kappa\beta_{\alpha_j}\ph(x_j)$ combined with the term
\be \int\!\!d^dx_id^dx_j\, \frac{\delta}{\delta h_{\mu\nu}(x_i)}\langle h_{\mu\nu}(x_i)\, \ph(x_j)\rangle\frac{\delta}{\delta\ph(x_j)}\ \in\ \Po_{ij}\,,\ee
which appears inside the exponential $\mathrm{e}^{\Po_{ij}}$ as part of the $\Gamma_n$ expansion, results in a shift operator for $h_{\mu\nu}$ that maps 
\be\label{hshift} h_{\mu\nu}(x_i) \mapsto h_{\mu\nu}(x_i)+\kappa\beta_{\alpha_j}\langle h_{\mu\nu}(x_i)\, \ph(x_j)\rangle\,,\ee
in the $i^\mathrm{th}$ copy of $\Gamma_1$. Using  \eqref{hplim}, this is
\be\label{hshiftexp} h_{\mu\nu}(x_i) \mapsto h_{\mu\nu}(x_i) + \kappa\beta_{\alpha_j}\left(\frac1{\varepsilon^2}-\frac{2d^2+\gamma}{8(d-1)(d-2)}+O(\varepsilon^2)\right) \Delta_{\mu\nu}(x_i-x_j)\,,\ee
where\footnote{The coefficient of $-\frac12$ can be found by applying $\partial_\mu$ to both sides and comparing the result to $\partial_\mu\Delta(x)$ using \eqref{prop}.}
\be\label{Deltamunu} \Delta_{\mu\nu}(x) = \int\!\!\frac{d^dp}{(2\pi)^d}\, \frac{\mathrm{e}^{ip\cdot x}}{\TRM{p^2}}  \left( \frac{\delta_{\mu\nu}}{d}-\frac{p_\mu p_\nu}{p^2}\right) = -\frac12 \left(\frac{\delta_{\mu\nu}}{d}-\frac{x_\mu x_\nu}{x^2}\right)\Delta(x)\,.\ee
Applying these shifts gives the final form of the resummed $\langle h_{\mu\nu}\ph\rangle$ corrections. 

One might worry that these shifts dominate in such a way as to destroy the regulating exponential \eqref{regulategen}. This depends on how we parametrise $\hat{g}_{\mu\nu}$, however if we use the exponential parametrisation \cite{Aida:1994zc,Nink:2014yya,Percacci:2015wwa}, which is a natural non-singular choice, then for all $d$ in the range $5/2<d<5$ it does not destroy the regularisation. This includes the important case of $d=4$ dimensions. 

To see this first note that in exponential parametrisation we write $\hat{g}_{\mu\nu} =(\mathrm{e}^{\kappa h})_{\mu\nu}$, where we are using the matrix exponential, the normalisation being fixed by \eqref{paramhg}. Then the exponential of $\Delta$ that appears as a result of the shift \eqref{hshiftexp} takes the form 
\be \exp\left\{ \kappa^2\hat{\beta}_i\beta_{\alpha_j}\left(\frac1{\varepsilon^2}-\frac{2d^2+\gamma}{8(d-1)(d-2)}+O(\varepsilon^2)\right)\Delta(x_i-x_j)\right\}\,,\ee
where the $\hat{\beta}_i$ are determined by the overall power $m_i$ of $\hat{g}_{\mu\nu}$ that appears in the corresponding interaction. \TRM{Now recall, \cf \eqref{pplim}, that $\xi^2=1/{\varepsilon^2}+O(\varepsilon^2)$ and thus the term in round brackets above is less than $\xi^2$.} For the given range of $d$, and by inspection, in all cases $\hat{\beta}_i<\beta_{\alpha_i}$, and thus this exponential never overwhelms the regularising exponential \eqref{regulategen}. Indeed, since $\hat{g}_{\mu\nu}$  has unit determinant,  one sees for example that in the Einstein-Hilbert action \eqref{EHreparam} the overall power of $\hat{g}_{\mu\nu}$ is $m=-1$, whereas for the gauge fixing and ghost actions in \eqref{gengf} it is $m=+2$ and $m=+1$ respectively. Now, from \eqref{Deltamunu},  $\Delta_{\mu\nu}$ is a sum of transverse and longitudinal projectors $\delta_{\mu\nu}-\frac{x_\mu x_\nu}{x^2}$ and $\frac{x_\mu x_\nu}{x^2}$, respectively, with corresponding eigenvalues $-\Delta/2d$ and $(d-1)\Delta/2d$. Thus the corresponding $\hat{\beta}=-m/2d$ and $(d-1)m/2d$. These should be compared to the corresponding $\beta_{\alpha_i}$ which in these examples are $(d-2)/d$, $4/d$ and $2/d$, for the Einstein-Hilbert action, gauge fixing and ghost actions respectively. One then readily verifies that in these cases, $\hat{\beta}_i<\beta_{\alpha_i}$ provided $5/2<d<5$. The cosmological constant and matter action cases are similarly verified. 

To this can we now add the other quantum corrections involving pure $h_{\mu\nu}$ propagators, the ghosts, and matter fields. Since there are interactions to all orders in $\kappa h_{\mu\nu}$, and there are $\half d(d+1)$ $h_{\mu\nu}$ versus just one $\ph$, and given that the graviton has a right-sign propagator, one might have thought that summing also over the pure $h_{\mu\nu}$ corrections before performing the integrals, must result in contributions that destroy the regularising properties of the $\ph$ propagator exponential (corrected as discussed above). In fact this limit shows that such an argument is too na\"\i ve: in this limit the $\langle h_{\mu\nu} h_{\alpha\beta} \rangle$ propagator \eqref{hhlim} remains finite, whilst the $\ph$ propagator corrections diverge. 

The size of the corrections can be studied by making the substitution 
\be \label{subs} x_i\mapsto \varepsilon^{-\frac2{d-2}}\,x_i\ee 
for the $x_i$ integration variable. This eliminates $\varepsilon$ from the regularising exponentials, renders finite any propagator factors involving ghosts or $\ph$, and forces all the $\langle h_{\mu\nu} h_{\alpha\beta} \rangle$ and matter field propagator corrections to vanish in the limit $\varepsilon\to0$. Furthermore the effect of the substitution \eqref{subs} in the Fourier exponential $\mathrm{e}^{ip\cdot x}$, is to take the large external momentum limit, which further suppresses the corrections. Unfortunately the substitution \eqref{subs} also supplies a divergent factor of $\varepsilon$ from the change of integration measure $\int\!d^dx_i$, and thus the $\Gamma_n$ get multiplied by a factor of $\varepsilon^{-\frac{(n-1)d}{d-2}}$. This means that increasing $n$ results in divergently larger corrections, and thus again we do not have a controlled expansion.

\section{Summary and Conclusions}
\label{sec:conclusions}

We have seen that the perturbation series in $\kappa$ can be resummed into terms, $\Gamma_n$, each of which is UV finite (provided that fields are parametrised appropriately, and massive tadpole corrections are set to zero).  These resummations into the $\Gamma_n$, involve an infinite rearrangement of the perturbation series in $\kappa$, since here we first expand over vertices in $S_I$ which keep intact the exponential dependence on $\ph$, and then sum over the $\ph$ corrections, before summing over  corrections from other fields. 

Such infinite rearrangements of a series can alter its value, unless that series is absolutely convergent. But the perturbation series in $\kappa$ is undoubtedly not even conditionally convergent, if only because the resulting Feynman diagrams suffer from the usual factorial growth.\footnote{For a discussion of such factorial growth, and Borel resummation \etc see \eg ref. \cite{ZinnJustin:2002ru}.} Had the $\Gamma_n$ obeyed all the required properties of such a quantum field theory, we could try to  take the view that they \emph{define} what we mean by quantum gravity. But the fact that no finite resummation of the $\Gamma_n$ is diffeomorphism invariant, either under background diffeomorphisms or its BRST realisation, obstructs any such attempt, as does the related problem: the apparent lack of any appropriate small parameter which would allow the $\Gamma_n$ to be ordered into successively smaller terms.

\TRM{As already mentioned in the Introduction, sec. \ref{sec:Introduction}, the idea that gravity should somehow provide its own UV regularisation is not new. In fact ref. \cite{Isham:1972pf} suggested an approach to gravity regularisation of quantum electrodynamics which is in some ways close  to this paper. They use an exponential parametrisation of the metric. As a model approximation, they retain only the conformal factor $\ph$ part of the metric. Working in Feynman -- DeDonder gauge, they note that summation over all the $\ph$ propagator corrections then results  in an exponential  regularisation factor for the simplest quantum corrections, the same exponential regularisation as derived  here, \viz eqn. \eqref{regulategen}.\footnote{\TRM{In pure Yang-Mills theory, summation over those classes of diagrams gives charge screening \cite{Khriplovich:1969aa},  understood eventually to be part of the asymptotically free one-loop running of its coupling.}}  That gravity might be regularised by resummation was also suggested in ref. \cite{DeWitt:1964yh}, although there the approach advocated was resummation of ladder diagrams in the Bethe-Salpeter equations. A different key idea used here was central to ref. \cite{tHooft:2010xlr}, namely that the conformal factor should be integrated out first. There, it was then argued that conformal invariance would imply that the result has to be UV finite.}

We hope that future work finds a way to build \TRM{on the findings we have reported, to achieve a fully} acceptable finite quantum field theory. Perhaps one can make further progress by first studying simpler models. For example the scalar field theory defined by the action 
\be S= \int\!\! d^dx \, \left\{-\frac12 (\partial_\mu\ph)^2 + \mu^d \,\mathrm{e}^{\kappa\ph}\right\}\,,\ee
or equivalently with a right sign kinetic term and $\mathrm{e}^{i\kappa\ph}$ interaction (see also the end of sec. \ref{sec:second}), 
can be quantised by these methods. Thus we see that it is UV finite in $d>2$ dimensions, and furthermore in this case the orders $\Gamma_n$ have a genuine expansion parameter, namely the mass parameter $\mu$ (although it can be changed by shifting $\ph$ by a constant). \TRM{It can be viewed as a kind of generalisation of Liouville field theory \cite{Distler:1988jt} to $d>2$ dimensions, albeit with the wrong sign kinetic term. Note that} such a theory however is non-unitary, unlike quantum gravity.

In summary, we have established that quantum gravity can be resummed into a series of UV finite terms $\Gamma_n$, but the apparent lack of diffeomorphism invariance in these $\Gamma_n$, or useful control parameter, is a serious stumbling block. Nevertheless this  result seems significant.

\vskip3cm

\section*{Acknowledgements}
\TRM{The author acknowledges support from STFC through Consolidated Grant ST/T000775/1.}

\clearpage

\appendix 

\section{Gauge fixing Yang-Mills fields}
\label{app:YM}

\TRM{In sec. \ref{sec:two} we showed that all interactions in a full phenomenologically relevant theory of quantum gravity can satisfy the key property that they are weighted by a positive exponential of $\ph$. However we left to this appendix the task to show that this is true also of Yang-Mills gauge fixing terms and corresponding ghost action.

The result of gauge fixing is to add to the Yang-Mills action \eqref{A}, the gauge fixing and ghost terms, which take the form
\be\label{YMgf} 2\sqrt{g}\, \mathrm{tr}\left( \chi f^2/2 - \bar{\eta} qf\right)\,,\ee
where $\chi$ is the gauge parameter, the gauge field $A_\mu = A^a_\mu t^a$ and ghosts $\eta = \eta^a t^a$ are  contracted into the generators $t^a$ of the Lie group (with conventional orthonormalisation tr$\,t^a t^b = \delta^{ab}/2$), $f$ is a suitable gauge fixing functional, typically $f= g^{\mu\nu}\nabla_\mu A_\nu$, and $q$ is the Yang-Mills BRST charge:
\be\label{YMBRS} q A_\mu = \nabla_\mu \eta -i \textrm{g} [A_\mu,\eta]\,,\qquad q \eta = -i \textrm{g} \eta^2\,,\qquad q\bar{\eta}=\chi f\,,\ee
with $\mathrm{g}$ being the Yang-Mills coupling. 

Unlike the Yang-Mills action \eqref{A}, the gauge fixing and ghost terms in \eqref{YMgf} are already weighted by a positive exponential of $\ph$ in the phenomenologically relevant case of $d=4$ dimensions (and in general $d$ as $\mathrm{e}^{\kappa(d-2)\ph/d}$ and $\mathrm{e}^{\kappa (d-1)\ph/d}$ respectively). The repairs to $A_\mu$ put forward in sec. \ref{sec:two} only make this better. 

Writing $A^\mu$ as the fundamental field removes the inverse metric from the gauge fixing functional, which now becomes $f=\nabla_\mu A^\mu$, and thus leaves us just with the same weight as a cosmological constant term, \cf \eqref{cc}, and making this substitution in \eqref{YMBRS} simply has the net effect of raising the indices in the first equation: $q A^\mu = \nabla^\mu \eta -i \textrm{g} [A^\mu,\eta]$. 

The other choice of repair replaces $A_\mu$ everywhere it appears with $g^{w/2}A_\mu$ and thus also only improves \eqref{YMgf}, whilst altering the  BRST transform to 
\be q A_\mu = g^{-w/2} \nabla_\mu \eta -i \textrm{g} [A_\mu,\eta]\,.\ee 

Since the metric plays no r\^ole in the Yang-Mills BRST symmetry, and commutes with the covariant derivative $\nabla_\alpha$, it should be clear that such changes of field variable do not break the Yang-Mills BRST symmetry and in particular the (on-shell) nilpotency of $q$, but it is also straightforward to confirm this from the above equations. 

(In the latter case we can return $qA_\mu$ to its standard form by replacing the ghost field $\eta$ with $g^{w/2} \eta$ and thus declaring it to be also a density of weight $w$. This only increases the positivity of the $\ph$ exponentials still further in the ghost action. Although this change to $\eta$ may thus seem motivated, it does not eliminate $g^{w/2}$ from the BRST algebra since it now appears in $q\eta$.)}

\bibliographystyle{hunsrt}
\bibliography{references}


\end{document}